%% file: lbs_2018_journal_indepenent.tex
\documentclass[]{interact}

\usepackage{epstopdf}
\usepackage{graphicx}        
\graphicspath{{../}}

\usepackage{natbib}
\bibpunct[, ]{(}{)}{;}{a}{}{,}
\renewcommand\bibfont{\fontsize{10}{12}\selectfont}

\theoremstyle{plain}

\theoremstyle{definition}

\theoremstyle{remark}

\usepackage{amsmath}
\usepackage{amsfonts}
\usepackage{amssymb}
\usepackage{mathptmx}
\usepackage{txfonts} 
\DeclareMathAlphabet{\mathpzc}{OT1}{pzc}{m}{it} 
\usepackage{bbm,bbold}

\usepackage{makeidx}
\usepackage{dcolumn}

\usepackage{multicol}        
\usepackage{acronym}
\usepackage{multirow}
\usepackage{color, soul}

\usepackage[caption=false, font=footnotesize]{subfig}
\usepackage{setspace}

\usepackage{footnote}
\makesavenoteenv{tabular}
\makesavenoteenv{table}
\usepackage{lineno}

\makeindex 

\begin{document}

\input{cz_acronyms.tex}
\input{cz_command.tex}

\articletype{Research paper}

\title{Modified Jaccard Index Analysis and \colorRev{Adaptive}  Feature Selection for Location Fingerprinting with Limited Computational Complexity}

\author{
\name{Caifa Zhou\textsuperscript{a}\thanks{This paper is a significantly revised and extended version of \citep{Zhou2018}, a paper published in the proceedings of LBS 2018, Zurich.} \thanks{CONTACT Caifa Zhou. Email: caifa.zhou@geod.baug.ethz.ch. (ORCID: 0000-0002-3304-5497)} and Andreas Wieser\textsuperscript{a}}
\affil{\textsuperscript{a}Institute of Geodesy and Photogrammetry, ETH Zurich, 8093, Zurich, Switzerland}
}

\maketitle

\begin{abstract}
	We propose an approach for \colorText{\acl{fbp}} which reduces the data requirements and computational complexity of the online positioning stage. It is based on a segmentation of the entire \colorText{\acl{roi}} into subregions, identification of candidate subregions during the online-stage, and position estimation using a preselected subset of relevant features. The subregion selection uses a \colorText{\acl{mji}} which quantifies the similarity between the features observed by the user and those available within the \colorText{\acl{rfm}}. The \colorRev{adaptive} feature selection is achieved using an \colorText{\acl{foba}} which determines a subset of features for each subregion, relevant with respect to a given \colorText{\acl{fbp}} method. \colorText{In an empirical study using signals of opportunity for fingerprinting the proposed subregion and feature selection reduce the processing time during the online-stage by a factor of about 10 while the positioning accuracy does not deteriorate significantly. In fact, in one of the two study cases the $ \texSym{90^{th}} $ percentile of the circular error increased by 7.5\% while in the other study case we even found a reduction of the corresponding circular error by 30\%.}
\end{abstract}


\section{Introduction}
\label{sec:1}
Fingerprinting-based indoor positioning systems (\acsp{fips}) are attractive for providing location of users or mobile assets because they can exploit \acf{sop} and infrastructure already existing for other purposes \citep{he2016wi}. They require no or little extra hardware, \citep{he2016wi}, and differ in that respect from many other approaches to indoor positioning like e.g., the ones using infrared beacons \citep{1432143}, ultrasonic signals \citep{hazas2006broadband}, \acf{ble} beacons \citep{7915011}, \acf{rfid} tags \citep{bekkali2007rfid}, \acf{uwb} signals \citep{ingram2004ultrawideband}, or foot-mounted \acfp{imu} \citep{Gu2017}. \acs{fips} benefit from the spatial variability of a wide variety of observable features or signals like \acf{rss} from \acf{wlan} \acfp{ap} \colorText{\citep{Padmanabhan2000, Youssef2008,7275492, Jun2017,Mendoza-Silva2018}}, magnetic field strengths \colorText{\citep{6860809, 7346763,Xie2016}}, or \acs{rss} of cellular towers \citep{7743656}. \colorText{Such signals are location dependent features, many of which can easily be measured using a variety of mobile devices (\eg smart phones or tablets).} \acsp{fips} are therefore also called feature-based indoor positioning systems \citep{6597797}. The attainable quality of the position estimation using \acs{fips} mainly depends on the spatial gradient of the features and on their stability or predictability over time \citep{Ndrmyr2014}.

Key challenges of \acs{fips}, especially the ones using \acs{rss} readings from \acs{wlan} \acsp{ap}, are discussed e.g., in \citep{kushki2007kernel} and more recently in \colorRev{\citep{he2016wi, yassin_recent_2017}}. The former publication focuses on four challenges of \acs{fips} utilizing vectors of \acs{rss} from \acs{wlan} \acs{ap} as features. In particular, the paper addresses i) the generation of a fingerprint database to provide a \acf{rfm} for positioning, ii) pre-processing of fingerprints for reducing computational complexity and enhancing accuracy, iii) selection of \acsp{ap} for positioning, and iv) estimation of the distance between a fingerprint measured by the user and the fingerprints represented within in the reference database. Extensions to large indoor regions and handling of variations of observable features caused by the changes of indoor environments or signal sources of the features (e.g., replacement of broken \acsp{ap}) are addressed in \citep{he2016wi}.

\colorText{Regarding generation of the \acs{rfm}, various approaches have been proposed \citep{he2016wi}. Initially, the features were mapped by dedicated surveying measurements (\eg \citep{Youssef2008,park2010growing}). The resulting \acsp{rfm} were accurate at the time of acquisition. However, such measurements are labor-intensive and time-consuming. Approaches based on forward modeling, e.g., using indoor propagation models \colorRev{\citep{6027190,El-Kafrawy2010,bisio_trainingless_2014}} or ray tracing \citep{8445928}, were proposed as a cost-effective alternative, especially in case of radio frequency signal strengths as features. However, the accuracy of the resulting \acs{rfm} depends on the validity of the model assumptions including the wave propagation models, the geometry and material properties of the objects and structures in the indoor space, and the location and antenna gain patterns of the signal sources. For many real-world application scenarios it is thus typically lower than the accuracy of measurement-based \acs{rfm} generation. More recently, approaches for \acs{rfm} generation using the sensors built into the mobile user devices have been proposed. They can be differentiated according to the degree of user participation. Data collection for \acs{rfm} generation can be done by an application running in the background such that the user only needs to consent to contributing data but not actively participate otherwise. Other approaches require the user to manually indicate his/her location on a map or to signal to the application that the current location corresponds to a certain marked ground truth, see e.g. \citep{6216368,6071942,6681046}. For refining the \acs{rfm} various approaches have been proposed, \eg interpolation/extrapolation \citep{7024940,6866898} or kernel-smoothing \citep{Huang2016}. We are not focusing on reducing the workload for building the \acs{rfm}. However, we validate the proposed approach using a kinematically collected \acs{rfm} (see \secPref\ref{sec:experimentalResults}) which is similar to a standard survey but less time-consuming.}


The \colorText{contribution} of this paper is to reduce the online positioning computation complexity by introducing \colorText{a specific approach to} subregion selection and feature selection into the online positioning process. \colorText{The proposed approach includes machine learning algorithms, i.e. algorithms whose performance on the specific task improves with experience \citep{Mitchell:1997:ML:541177} and which are thus suitable for \acl{fbp} \citep{Bishop:2006:PRM:1162264}.} \colorText{It} is applicable to \acsp{fips} using opportunistically measured location-relevant features. In case of \acsp{fips} using \acs{sop}, the difficulty lies in both the type and number of features \colorText{varying across the region of interest.} This introduces a critical limitation which prevents the applicability of the aforementioned \acs{fips} solutions due to changes of the dimension of the feature space across time and across location coordinate space. For instance, the \acl{fbp} methods including the typical ones, \eg $ k $-nearest neighbors ($ k $NN) \citep{Padmanabhan2000} and \acf{map} \citep{Youssef2008}, and the advanced ones, \eg \acl{fbp} using \acf{svm} \citep{1297088}, \acf{lda} \citep{Nuno-Barrau2006}, Bayesian network \citep{Nandakumar:2012:CLD:2348543.2348579}, and Gaussian process \citep{6018966}, cannot be applied without special precautions for handling the varying dimension in such cases. \colorText{Few previous publications address handling this problem \eg \citep{ZhouIpin2018}}. Additionally, the computational complexity of these \acl{fbp} methods is proportional to the number of reference locations in the \acs{rfm} and the number of observable features. \colorText{This makes these approaches computationally expensive in large \acsp{roi} with many reference locations and many available features unless introducing specific means for mitigation.} The discrepancy of the feature dimension and the computational complexity problems is typically mitigated by introducing subregion selection and feature selection \citep{Zhou2018,kushki2007kernel,feng2012received,6018966,7752987}.

\colorText{In this paper we propose (i) subregion selection based on a modified Jaccard index (MJI), (ii) a forward-greedy search to find an appropriate number of subregions, and (iii) an adaptive forward-backward greedy search  (\colorRev{\acs{foba}}) algorithm \citep{zhang2011adaptive} for selecting the relevant features for each subregion. We demonstrate the application of the proposed algorithms to both MAP- and kNN-based position estimation. We finally validate the performance of the proposed approach by carrying out experiments in two RoIs of different size using two types of opportunistically measured signals (i.e., WLAN and BLE).}

The structure of the paper is as follows: A short review of the previous publications addressing the approaches for subregion selection and feature selection is given in \secPref\ref{sec:related}. In \secPref\ref{sec:proposedApproach}, the \acs{mji}-based subregion selection, \colorRev{\acs{foba}}-based feature selection, and the modified online positioning process are presented along with the computational complexity. We illustrate and validate the performance of the proposed approach in \secPref\ref{sec:experimentalResults} by applying it to an \acs{fips}, which utilizes \acs{sop} as the location-relevant features, and we compare the results to those obtained using previously proposed methods. 

\begin{table}[!htb]
	\centering
	\colorText{
		\caption{Selected acronyms used herein}
		\label{tab:acronyms}
		\begin{tabular}{p{0.1\columnwidth}p{0.55\columnwidth}}
			\hline
			Acronym & Meaning\\
			\hline
			\acs{fips}&\acl{fips}\\
			\acs{rss}&\acl{rss}\\
			\acs{sop}&\acl{sop}\\
			\acs{rfm}&\acl{rfm}\\
			\acs{knn}&\acl{knn}\\
			\acs{map}&\acl{map}\\
			\acs{mji}&\acl{mji}\\
			\colorRev{\acs{foba}} &\acl{foba}\\
			\acs{lasso}&\acl{lasso}\\
			\acs{mse}&\acl{mse}\\
			\acs{ecdf}&\acl{ecdf}\\
			\hline
	\end{tabular}}
\end{table}

\section{Related work}\label{sec:related}

\subsection{Subregion selection}\label{subsec:subregion}

The subregion selection process contributes to constraining the search space. The selected subregions are treated as coarse approximations of the user's location. The process of refining the coordinate estimates is then carried out only within these subregions and the search space \colorText{for the final estimate} is thus independent of the size of the \acs{roi}. There are mainly two types of approaches for subregion selection\footnote{In other publications, subregion selection is called spatial filtering \citep{kushki2007kernel}, location-clustering \citep{youssef2003wlan}, or coarse localization \citep{feng2012received}.}: approaches based on clustering and approaches based on similarity metrics. \citep{feng2012received, Karegar2017}, and \citep{chen2006power, 6018966} applied affinity propagation and $k$-means clustering to divide the \acs{roi} into a given number of subregions according to the features collected within the \acs{roi}. Both papers present clustering-based subregion selection\colorText{, }and require prior definition of the desired number of subregions and knowledge of all features observable within the entire \acs{roi}. These clustering-based approaches take the fingerprint measured by the user into account during the clustering process which thus has to be repeated with each new user fingerprint obtained.

Similarity metric-based subregion selection relies on the identification of the subregions whose fingerprints contained in the \acs{rfm} are most similar to the fingerprint observed by the user. They differ depending on the chosen similarity metric. E.g., \citep{kushki2007kernel} use the Hamming distance for this purpose, measuring only the difference in terms of observability of the features, not their actual values. Still, these approaches typically need prior information on all observable features within the entire \acs{roi} when associating a user observed features with a subregion. This may be a severe limitation in case of a large \acs{roi} or changes of availability of the features. \acs{mji}-based subregion selection as \colorText{proposed} in this paper belongs to similarity metric-based subregion selection. However, the approach proposed herein requires only the prior knowledge of the features observable within each subregion when \colorText{quantifying} the similarity metric between the observations in the \acs{rfm} and in the user observed measurements.

\subsection{Selection of relevant features}\label{subsec:figerprintSelection}

Approaches to selection of features \colorText{(or sparse representation)} actually used for positioning differ \acs{wrt} several perspectives. We focus on three aspects in their review: i) whether they take the relationship between positioning accuracy and selected features into account, ii) whether they help to reduce the computational complexity of position estimation, and iii) whether they are applicable to a variety of features or only features of a certain type. The chosen features for positioning should be the ones allowing to achieve the best positioning accuracy using the specific \colorText{\acl{fbp}} method or achieving a useful compromise between accuracy and reduced computational burden. 

Previous publications focused on feature selection for \acs{fips} using \acs{rss} from \acs{wlan} \acsp{ap} and consequently addressed the specific problem of \acs{ap} selection rather than the more general feature selection. \citep{chen2006power} and \citep{feng2012received} proposed using the subsets of \acsp{ap} whose \acs{rss} readings are the strongest assuming that the strongest signals provide the highest probability of coverage over time and the highest accuracy. \citep{kushki2007kernel} and \citep{chen2006power} applied a divergence metric (Bhattacharyya distance and information gain, respectively) to minimize the redundancy and maximize the information gained from the selected \acsp{ap}. The limitations of these approaches are: i) they are only applicable to the \acs{fips} based on \acs{rss} from \acs{wlan} \acsp{ap}, and ii) they only take the values of the features into account as selection criteria instead of the actual positioning accuracy. \citep{5210101} proposed an \acs{ap} selection strategy able to choose \acsp{ap} ensuring a certain positioning accuracy using a nonparametric information filter. However, this approach uses consecutively measured fingerprints to select the subset of \acsp{ap} maximizing the discriminative ability \acs{wrt} localization. This method therefore needs several online observations for estimating one current position.

In \citep{Zhou2018}, a feature selection algorithm based on randomized \acs{lasso} \citep{tibshirani1996regression}, an $ L_1 $-regularized linear regression model, for selecting the relevant features for positioning is proposed. Each feature within the subregion is associated with an estimated coefficient. If the coefficient is sufficiently different from zero the corresponding feature is identified as relevant. However, this approach only connects the feature selection with the positioning error indirectly. \colorText{Furthermore}, \acs{lasso}-based feature selection is equivalent to \acs{map} \colorText{if} the likelihood is Gaussian and the prior distribution is Laplace \citep{doi:10.1198/016214508000000337}. These two assumptions are not 
necessarily justified with \colorText{\acl{fbp}}. It is difficult to find a proper value of the hyper-parameter of \acs{lasso}-based feature selection, which makes the feature selection unstable \citep{fastrich2015constructing}. Finally, this feature selection algorithm is prone to overfitting. Applying this approach to select the relevant features for each subregion requires the number of observations in each subregion to be much larger than the number of the dimension of the features (e.g., the number of observable \acs{wlan} \acsp{ap} or \acs{ble} beacons). Normally, this requirement is not met in case of \colorText{\acl{fbp}} using \acsp{sop} as the features.

In this paper, we thus propose an approach based on \colorRev{\acs{foba}} to choose the most relevant features for \colorText{\acl{fbp}}. This method differs from the previously mentioned ones in three ways: i) it takes the positioning error into account directly, i.e. the feature selecting process is directly combined with the \colorText{\acl{fbp}} methods that are used at the online stage, ii) wrongly selected features from the forward greedy search step can be adaptively corrected  by a backward greedy search step \citep{zhang2011adaptive}, and iii) it is a data-driven algorithm adapting automatically to the number of observations of each subregion.

\section{The proposed approach}\label{sec:proposedApproach}

In this section, we briefly summarize the fundamentals of \colorText{\acl{fbp}} and present the main contributions of this paper to reduce the computational complexity independent of the size of the \acs{roi}. In particular we present  i) candidate subregion selection according to \acs{mji}, ii) selection of relevant features using \colorRev{\acs{foba}}, and iii) adaptations of \acs{map} and \acs{knn}-based positioning with the combination of the previous two steps. Finally we briefly discuss the computational complexity of the proposed method.

\subsection{Problem formulation}\label{subsec:problemFormulation}



\colorText{Each measured feature has a unique identifier and a measured value, \eg the measurement related to a specific WiFi \acs{ap} can be identified by the \acf{mac} address and has a \acs{rss}. It is thus formulated as a pair $ (a, v) $ of attribute $ a $ and value $ v $. A complete measurement (\ie fingerprint) $ \setSymScript{O}{i}{u} $ taken by user $ \mathrm{u} $ at location/time $ i $ consists of a set of attribute-value pairs, \ie $ \setSymScript{O}{i}{u} := \{(\norSymScript{a}{ik}{u}, \norSymScript{v}{ik}{u}) | \norSymScript{a}{ik}{u}\in\setSym{A}; \norSymScript{v}{ik}{u}\in\convSetSym{R};k\in\intSet{\norSymScript{N}{i}{u}} \} $,  where $ \setSym{A} $ is the complete set of the identifiers of all available features and $ N_i^{\texSym{u}} $ ($ N_i^{\texSym{u}} = |\setSymScript{O}{i}{u}|$) is the number of features observed by $ \texSym{u} $ at $ i $. The set of keys of such a fingerprint is defined as $ \setSymScript{A}{i}{u} := \{a_{ik}^{\texSym{u}}|\exists (a_{ik} ^{\texSym{u}}, v_{ik})\in \setSymScript{O}{i}{u}\} $.  A discrete \acs{rfm} $ \setSym{M} := \{(\vecSym{l}_j, \setSymScript{\tilde{O}}{j}{})| \vecSym{l}_j\in \setSym{G}, j\in\intSet{|\setSym{M}|} \} $ is given as a set of position-fingerprint pairs representing the relationship between the location $ \vecSym{l} $ and the features $ \setSym{O} $ at different locations within the \acs{roi} $ \setSym{G} $. } 

\colorText{\colorRev{Specifically for the subregion selection, $ \setSym{G} $ is divided into $ M $ non-overlapping subregions of arbitrary shape, \ie $ \setSym{G} = \underset{i=1:M}{\cup}\,\setSym{g}_i,\,\setSym{g}_i \cap \setSym{g}_j=\emptyset | i\ne j $. This segmentation can take contextual information into account by defining the subregions such that one subregion lies only in e.g. one building, one floor, one room or one corridor. Thus the concept is directly applicable to multi-building or mutli-floor situations.} Each of the measurements (elements) of $ \setSym{M} $ is assigned to the corresponding subregion. The subset of $ \setSym{M} $ corresponding to the $ g^{\texSym{th}} $ subregion can thus be defined as $ \setSymScript{M}{g}{} := \{(\vecSym{l}_j, \setSymScript{\tilde{O}}{j}{})|\vecSym{l}_j\in\setSymScript{g}{g}{},j\in\intSet{|\setSymScript{M}{g}{}|}\}, \setSymScript{M}{g}{}\subseteq \setSym{M}$, and the set of observable features in the corresponding subregion is denoted as $ \setSymScript{A}{g}{} := \{a|a\in\underset{j=1:|\setSymScript{M}{g}{}|}{\cup}{\setSymScript{\tilde{A}}{j}{}},\setSymScript{\tilde{A}}{j}\,\text{ of } \setSymScript{\tilde{O}}{j}{}, \exists (\vecSym{l}_j, \setSymScript{\tilde{O}}{j}{})\in\setSymScript{M}{g}{}\} (g\in\intSet{M}), \setSymScript{A}{g}{}\subseteq \setSym{A}$.}

\colorText{The positioning process consists of inferring the estimated user location $ \vecSymScript{\hat{l}}{i}{u}=\funSym{f}(\setSymScript{O}{i}{u}) $ as a function of the fingerprint and the \acs{rfm} where $\funSym{f}$ is a suitable mapping from fingerprint to location, \ie $ \funSym{f}:\setSymScript{O}{i}{u}\mapSym \vecSymScript{\hat{l}}{i}{u}$. Herein we propose the following solutions for mitigating the computational load associated with the online stage:}
\begin{itemize}
	\item identifying \colorText{(during the offline stage)} the most relevant features within each \colorText{subregion} using the \colorRev{\acs{foba}} such that the actual location calculation can be carried out \colorText{during the online stage} using only those instead of using all features;
	\item selecting the subregion as a coarse approximation of the actual user location based on \acs{mji} during the online-stage; 
\end{itemize}

Within this paper we combine the above two steps with a \acs{map} and \acs{knn}-based positioning \colorText{for performance analysis. We implement }it in a way to keep the computational complexity of the online stage almost independent of the size of the \acs{roi} and of the total number of observable features within the \acs{roi}.
\begin{figure}[!htb]
	\centering
	\includegraphics[width=0.75\columnwidth]{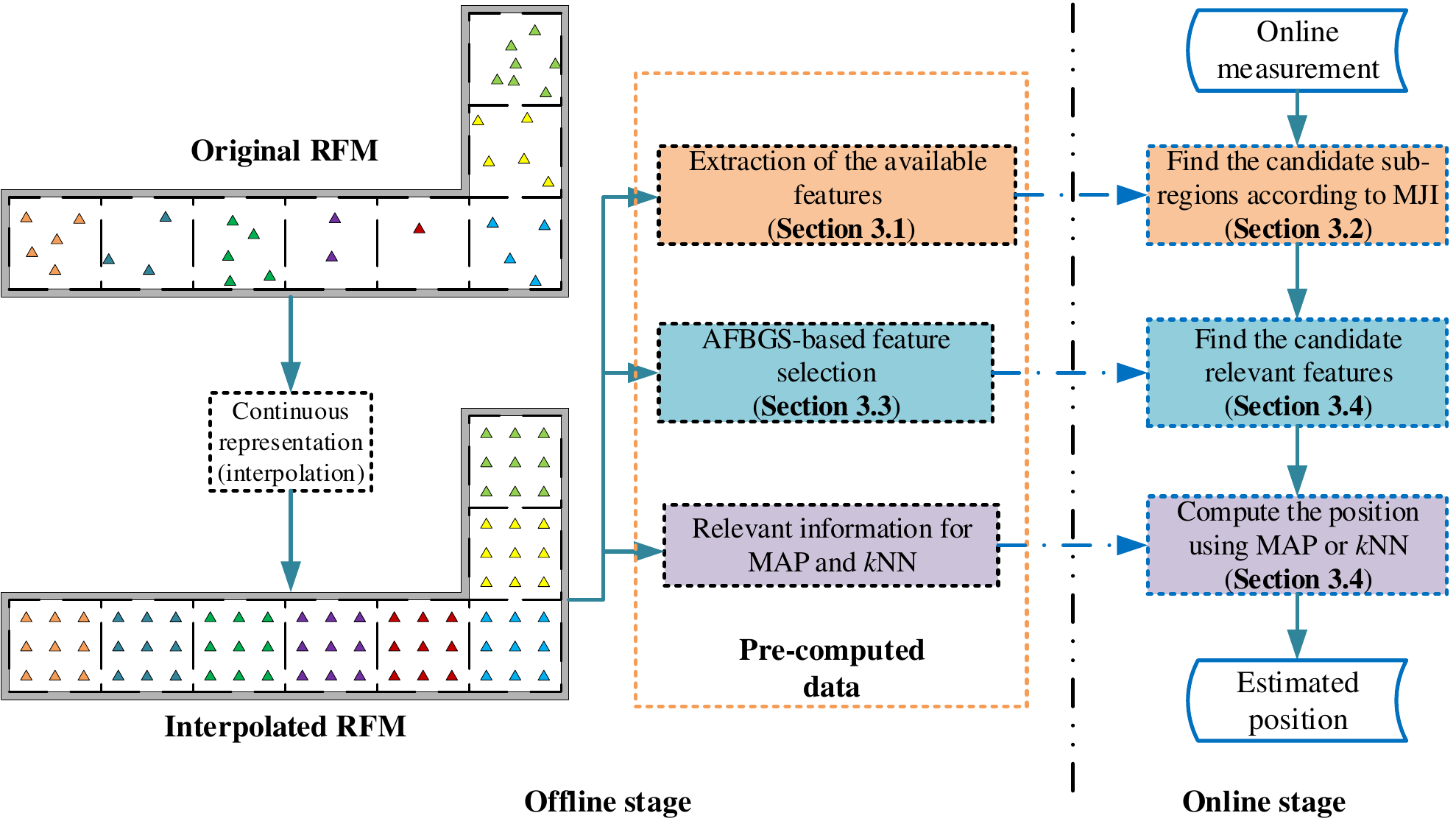}
	\caption{The proposed framework. In order to make the number of reference points within all subregions equal we interpolate the reference data in the original \acs{rfm} \colorText{using kernel smoothing} \citep{Berlinet2011} to provide a denser regular grid of reference points. This interpolated \acs{rfm} is used to calculate the pre-computed data for online positioning.}
	\label{fig:schematic}
\end{figure}

\subsection{Subregion selection using \acs{mji}}\label{subsec:modifedJin}
\begin{figure}[!htb]
	\centering
	\colorText{
	\subfloat[High \acs{mji} value]{
		\label{subfig:mji_high}
		\includegraphics[width=0.18\columnwidth]{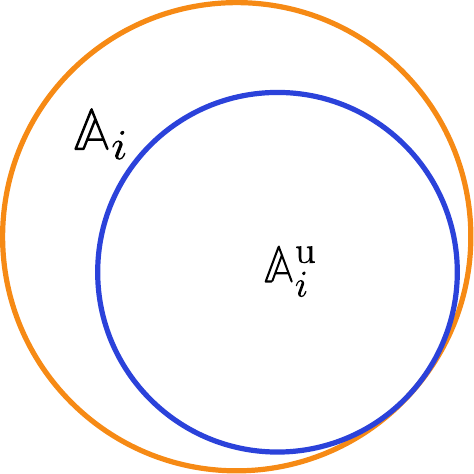}}\hspace{2ex}
	\subfloat[High \acs{mji} value]{
		\label{subfig:mji_inverse}
		\includegraphics[width=0.18\columnwidth]{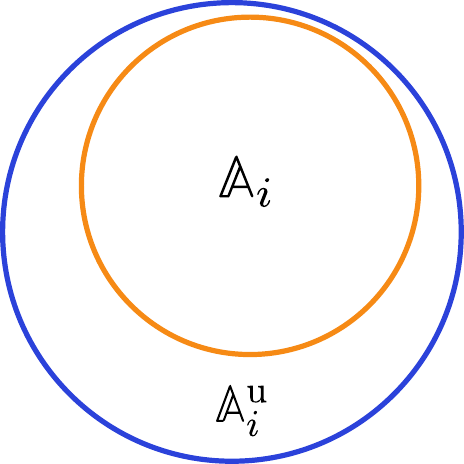}}\hspace{2ex}
	\subfloat[Low \acs{mji} value]{
		\label{subfig:mji_low}
		\includegraphics[width=0.18\columnwidth]{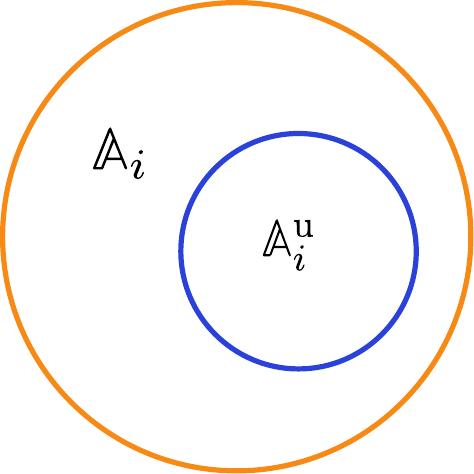}}\hspace{2ex}
	\subfloat[Low \acs{mji} value]{
		\label{subfig:mji_intersect}
		\includegraphics[width=0.253\columnwidth]{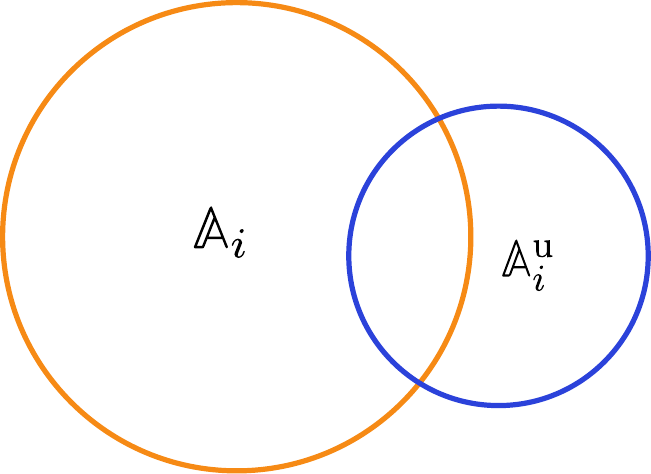}}
	\caption{Qualitative examples of \acs{mji} values: (a) a few features within the current subset of the \acs{rfm} are not measured, (b) there are some measured features which are not in the current subset of the \acs{rfm}, (c) the measured features are only few of the ones in the current subset of the \acs{rfm}, (d) there is little overlap between the measured features and the ones in the current subset of the \acs{rfm}.}
	\label{fig:mjiExamples}}
\end{figure}

\acs{mji}, an indicator of similarity between the keys of the measured fingerprints and the keys associated with the individual subregions in the \acs{rfm}, is applied to identify a set of candidate subregions most probably containing the actual user location \citep{Zhou2018,Jani2015}. We depict four qualitative examples in \figPref\ref{fig:mjiExamples}. If the overlap between the features within a  subregion (according to the \acs{rfm}) and the features within a user's observation is large, the value of the \acs{mji} is high, otherwise it is low. For a more detailed discussion of \acs{mji} see \citep{Zhou2018}.

The $ m $ subregions with the highest values of the \acs{mji} are selected as candidate subregions for the subsequent positioning. Their cell indices are collected in the vector $ \mathbf{s}^{\mathrm{u}}_m \in\mathbb{N}^m$ for further processing. If the subregions are non-overlapping, as introduced above, $m$ needs to be large enough to accommodate situations where the actual user location is close to the border between subregions and small enough to reduce the computational burden of the subsequent user location estimation.

In order to determine a proper value of $ m $, an indicator \colorText{is introduced} for denoting whether a validation example is contained in the selected top $ m $ subregions \colorText{ (indicator 1) or not  (indicator 0)}. \colorRev{Throughout this paper we assume that the user is actually within the RoI and thus within one subregion when the subregion selection process is carried out.} Given $ M^{\mathrm{val}} $ validation examples $ \{ (\setSym{O}_n^{\mathrm{val}}, \vecSym{l}_n^{\mathrm{val}})\}_{n=1}^{M^{\mathrm{val}}} $, the indicator for the $n$-th example is \colorText{thus} defined as:
\begin{equation}
	\label{eq:subregionAccuracyIndicator}
	I(\vecSym{l}_n^{\mathrm{val}}, m ) = \Bigg\{ 
	\begin{array}{ll}
		1, & \mathrm{if \,\,} \vecSym{l}_n^{\mathrm{val}} \in \underset{j\in\setSym{S}^{\vecSym{l}_n^{\mathrm{val}}}_m}\cup \setSymScript{g}{j}{} \\
		0, & \mathrm{otherwise}
	\end{array}
\end{equation}
where $  \setSym{S}^{\vecSym{l}_n^{\mathrm{val}}}_m $ contains the indices of the selected subregions for this particular validation example. \colorText{The subregion selection loss is then defined as the fraction of the validation examples for which the true location is not within the selected subregions,} \ie
\begin{equation}
	\label{eq:subregionSelectionLoss}
	\funSym{L}^{\mathrm{MJI}}(m) = 1 - \frac{\sum_{n=1}^{M^{\mathrm{val}}}I(\vecSym{l}_n^{\mathrm{val}}, m)}{M^{\mathrm{val}}}.
\end{equation}

\colorText{The subregion selection loss reduces as $m$ increases. However, increasing the number of selected subregions contradicts to the idea of reducing the computational complexity of the online positioning. Balancing between subregion selection loss and computational complexity is a multi-objective optimization problem, and it has no single optimum. Herein we decided to select an appropriate number of subregions heuristically by analyzing plots of selection loss versus $ m $ and choosing a value from which on the curve is flat. I.e., at which further increasing $ m $ hardly reduces the loss.}

\subsection{Feature selection using \colorRev{\acs{foba}}}\label{subsec:adaFoba}

In a real-world environment there may be a large number of features available for positioning, e.g., hundreds of \acsp{ap} may be visible to the mobile user device in certain locations. Not all of them will be necessary to estimate the user location. In fact using only a well selected subset of the available signals instead of all will reduce the computational burden and may provide a more accurate estimate because the measured signals are affected by noise and possibly interference. Furthermore the number of observable features typically varies across the \acs{roi} e.g., due to \acs{wlan} \acsp{ap} antenna gain patterns, structure and furniture of the building. However, it is preferable to use the same number of features throughout the candidate subregions for assessing the similarity between the measured fingerprint and the ones extracted from the \acs{rfm} during the online phase.

We therefore recommend selecting a number $h$ of features per candidate subregion for the final position estimation. To facilitate this selection during the online phase, the relevant features within each subregion are already identified beforehand once the \acs{rfm} is available. We use the \colorRev{\acs{foba}} \citep{zhang2011adaptive} for this step. During the online phase $h$ relevant features (possibly different for each subregion) are selected among the identified ones such that they are available both within the \acs{rfm} and the user fingerprint.

The combination of forward and backward steps within the algorithm\footnote{The forward part of the algorithm is referred to as matching pursuit for least square regression in the signal processing community \citep{10.2307/25464616, 258082}. In machine learning it is known as boosting \citep{buehlmann2006}.} makes \colorRev{\acs{foba}} capable of correcting intermediate erroneous selection of features. We first shortly review both forward and backward greedy algorithms, then present \colorRev{\acs{foba}}.


Let $ \setSym{F}_{(h_{i})} $ denote the set of indices of the selected relevant features of the $ i^{\mathrm{th}} $ subregion. A \colorText{\acl{fbp}} method $ \funSym{f }$ thus only uses these selected features to estimate the position, i.e. $\funSym{f}: \setSym{O} | \setSym{F}_{(h_{i})} \rightarrow\mathbf{l}$. The forward greedy search algorithm adds features one at a time picking the one causing the biggest reduction of a given loss at each step. \colorText{Herein we use the \acf{mse}, a widely applied loss function for regression problems,  as the loss metric for the feature selection.} For instance,  \colorText{with $ \setSym{F}_{(h_{i})} $ being the set of $ h $ features selected for the $i^{\mathrm{th}}$ subregion, the loss is the \acs{mse} of the positions estimated for all validation examples using this particular set of subregions:}
	\begin{equation}
	\label{eq:msewithfeatureselection}
	\funSym{L}^{\mathrm{FS}}(\setSym{F}_{(h_{i})}) = \frac{1}{M^{\mathrm{val}}}\sum_{n=1}^{M^{\mathrm{val}}}\|\funSym{f}(\setSymScript{O}{n}{val}|\setSym{F}_{(h_{i})}) - \mathbf{l}_{n}^{\mathrm{val}}\|_2^2
	\end{equation}
	An example of applying the forward greedy search to select features for the $ i^{\mathrm{th}} $ subregion is described in \tabPref\ref{tab:forward}. This algorithm works well in case the features are independent of each other. Otherwise the forward greedy search might wrongly select non-relevant features. Such wrong selections cannot be corrected by the forward greedy algorithm anymore. One remedy introduced to solve this problem is the backward greedy search (\tabPref\ref{tab:backward}), which starts with all features and greedily removes them one at a time picking the one associated with the smallest increase of the loss at each step. However, also the backward greedy search has two disadvantages: i) it is computationally costly because it starts with all features; ii) it is prone to overfitting if the number of observations in each subregion is much lower than the number of observable features. 
	
	A practically useful alternative is the \colorRev{\acs{foba}}, i.e., introducing a backward greedy search after each step of the forward greedy search. In this way the backward search starts with a set that does not overfit, and it can correct wrong selections made earlier.  Features removed by the backward steps are treated again as candidates during subsequent forward steps.

	\begin{table}[!h]
		\centering
		\caption{Pseudocode of the forward greedy search algorithm}
		\label{tab:forward}
		\begin{tabular}{p{0.01\columnwidth}p{0.89\columnwidth}}
			\hline
			& \textbf{Algorithm}: {forward greedy search}\\ 
			\hline
			1: & Input:$\,\,\,\{\setSymScript{O}{n}{val},  \vecSymScript{l}{n}{val} \}$, $ n=1,2,\cdots,M^{\mathrm{val}} $;\\
			& $\qquad\quad$minimum reduction of the loss $ \epsilon > 0$;\\
			& $\qquad\quad$maximum number of relevant features $ k \in \convSetSym{N}$;\\
			& $\qquad\quad$all observable features $\setSym{F}_{\mathrm{all}} $\\
			2: & Output: relevant features $ \setSym{F}_{(h_{i})} $ of $ i^{\mathrm{th}} $ subregion\\
			3: &  $ \setSym{F}_{(0)} = \emptyset$, initial \acs{mse} $ \funSym{L}^{\texSym{FS}}(\setSym{F}_{(0)}) $ \footnote{Herein the initial loss is defined as the \acs{mse} by using the median of all \acsp{rp} as the estimated position when no candidate feature is selected. }\\ 
			4: & \textbf{for} $ t =1,2,\cdots, |\setSym{A}_{i}| $:\\
			5: &$\quad$$ \setSym{F}_{\mathrm{avail}} = \setSym{F}_{\mathrm{all}}\backslash \setSym{F}_{(t-1)}$\\
			6: &$\quad$$ \hat{\setSym{F}}=\underset{\setSym{F}= \setSym{F}_{(t-1)} \cup \, \beta}{\arg\min}\funSym{L}^{\texSym{FS}}(\setSym{F})$, $\forall \beta \in \setSym{F}_{\mathrm{avail}}$\\
			7: & $\quad$$\setSym{F}_{(t)}=\hat{\setSym{F}}$\\
			8: & $\quad$$ \Delta = \funSym{L}^{\texSym{FS}}(\setSym{F}_{(t-1)}) -  \funSym{L}^{\texSym{FS}}(\setSym{F}_{(t)})$\\
			9: & $\quad$\textbf{if} ($\Delta \le \epsilon$ or $  |\setSym{F}_{(t)}| \ge k$):\\
			10: & $\quad$$\quad$ \textbf{break}\\
			11: & $\quad$\textbf{end if}\\
			12: & \textbf{end for}\\
			13: & \textbf{return} $ \setSym{F}_{(h_{i})} = \setSym{F}_{(t)}$: set of selected features\\
			\hline
		\end{tabular}
	\end{table}

	\begin{table}[!h]
		\centering
		\caption{Pseudocode of the backward greedy search algorithm}
		\label{tab:backward}
		\begin{tabular}{p{0.01\columnwidth}p{0.89\columnwidth}}
			\hline
			& \textbf{Algorithm}: {backward greedy search}\\ 
			\hline
			1: & Input:$\,\,\,\{\setSymScript{O}{n}{val},  \vecSymScript{l}{n}{val} \}$, $ n=1,2,\cdots,M^{\mathrm{val}} $;\\
			&$ \qquad\quad$maximum increment of the loss $ \phi > 0$; \\
			&$\qquad\quad$minimum number of relevant features $ k \in \convSetSym{N}$; \\
			&$ \qquad\quad$all observable features $\setSym{F}_{\mathrm{all}}  $\\
			2: & Output: relevant features $ \setSym{F}_{(h_{i})} $ of $ i^{\mathrm{th}} $ subregion\\
			3: & $ \setSym{F}_{(|\setSym{A}_{i}|)} =\setSym{F}_{\mathrm{all}}$ \\
			4: & \textbf{for} $ t =|\setSym{A}_{i}|-1, |\setSym{A}_{i}| - 2,\cdots, 1$:\\
			5: &$\quad$$ \hat{\setSym{F}}=\underset{\setSym{F}= \setSym{F}_{(t+1)} \backslash \beta}{\arg\min}\funSym{L}^{\texSym{FS}}(\setSym{F})$, $\forall \beta \in \setSym{F}_{(t+1)}$\\
			6: & $\quad$$\setSym{F}_{(t)}=\hat{\setSym{F}}$\\
			7: & $\quad$$ \Delta = \funSym{L}^{\texSym{FS}}(\setSym{F}_{(t)}) -  \funSym{L}^{\texSym{FS}}(\setSym{F}_{(t+1)})$\\
			8: & $\quad$\textbf{if} ($\Delta \ge \phi$ or $  |\setSym{F}_{(t)}| \le k$):\\
			9: & $\quad$$\quad$ \textbf{break}\\
			10: & $\quad$ \textbf{end if}\\
			11: & \textbf{end for}\\
			12: & \textbf{return} $  \setSym{F}_{(h_{i})} = \setSym{F}_{(t+1)}$: set of selected features\\
			\hline
		\end{tabular}
	\end{table}
	
	\tabPref\ref{tab:adaFoBa} shows the pseudocode of the implementation of \colorRev{\acs{foba}}. This algorithm \colorText{adaptively identifies} the relevant features for each subregion according to the chosen minimum reduction $ \epsilon $ and minimum relative increment $ \nu $ of the loss. The number of identified relevant features per subregion will therefore vary across the \acs{roi}. In \citep{zhang2011adaptive}, the author recommended to set $ \nu=0.5 $ which we do when applying the algorithm later on. {According to \citep{zhang2011adaptive}}, \colorRev{\acs{foba}} will terminate in a finite number of steps, which is no more than $ \lceil1 + \frac{\funSym{L}^{\texSym{FS}}(\setSym{F}_{(0)})}{\nu\epsilon}\rceil $, where $ \funSym{L}^{\texSym{FS}}(\setSym{F}_{(0)}) $ is the \acs{mse} {by taking the median of all \acsp{rp} as the initially guessed position}.
	 
	\begin{table}[!h]
		\centering
		\caption{Pseudocode of the \acf{foba} algorithm}
		\label{tab:adaFoBa}
		\begin{tabular}{p{0.01\columnwidth}p{0.89\columnwidth}}
			\hline
			& \textbf{{Algorithm}}: \colorRev{\acs{foba}}\\ 
			\hline
			1: & Input:$\,\,\,\{\setSymScript{O}{n}{val},  \vecSymScript{l}{n}{val} \}$, $ n=1,2,\cdots,M^{\mathrm{val}} $;\\
			&$ \qquad\quad$minimum reduction of the loss $ \epsilon >0$; \\
			&$\qquad\quad$relative increment of the loss $ \nu\in (0,1)$; \\
			&$ \qquad\quad $all observable features $\setSym{F}_{\mathrm{all}}  $\\
			2: & Output: relevant features $ \setSym{F}_{(h_{i})} $ of $ i^{\mathrm{th}} $ subregion\\
			3: & $t=1$, $ \setSym{F}_{(0)} = \emptyset$, initial \acs{mse} $ \funSym{L}^{\texSym{FS}}(\setSym{F}_{(0)})$ \\
			4: & \textbf{while (True)} \\
			5: &$\quad$$ \setSym{F}_{\mathrm{avail}} = \setSym{F}_{\mathrm{all}}\backslash \setSym{F}_{(t-1)}$\\
			6: &$\quad$$ \hat{\setSym{F}}^{\mathrm{forward}}=\underset{\setSym{F}= \setSym{F}_{(t-1)} \cup \,\beta}{\arg\min}\funSym{L}^{\texSym{FS}}(\setSym{F})$, $\forall \beta \in \setSym{F}_{\mathrm{avail}}$\\
			7: & $\quad$$\setSym{F}_{(t)}=\hat{\setSym{F}}^{\mathrm{forward}}$\\
			8: & $\quad$$ \Delta^{\mathrm{forward}} = \funSym{L}^{\texSym{FS}}(\setSym{F}_{(t - 1)}) -  \funSym{L}^{\texSym{FS}}(\setSym{F}_{(t)})$\\
			9: & $\quad$ \textbf{if} ($\Delta^{\mathrm{forward}} \le \epsilon$)\\
			10: & $\quad$$\quad$ \textbf{break}\\
			11: & $\quad$ \textbf{end if}\\
			12: & $\quad$$ \setSym{F}_{\mathrm{backward}}= \setSym{F}_{(t)}$\\
			13: & $\quad$$ c = |\setSym{F}_{\mathrm{backward}}| $, $ \setSym{F}_{(c)} =\setSym{F}_{\mathrm{backward}}$\\
			14:  & $\quad$ \textbf{while (True)} \\
			15: & $\quad$$\quad$ \textbf{if} ($ |\setSym{F}_{\mathrm{backward}}| $ == 1)\\
			16: & $\quad$$\quad$$\quad$ \textbf{break}\\
			17: & $\quad$$\quad$ \textbf{end if}\\
			18: &$\quad$$\quad$ $  c = c -1 $\\
			19: &$\quad$$\quad$ $ \hat{\setSym{F}}^{\mathrm{backward}}=\underset{\setSym{F}= \setSym{F}_{(c+1)} \backslash \beta}{\arg\min}\funSym{L}^{\texSym{FS}}(\setSym{F})$, $\forall \beta \in \setSym{F}_{(c+1)}$\\
			20: & $ \quad $$\quad$ $\setSym{F}_{(c)} = \hat{\setSym{F}}^{\mathrm{backward}}$\\
			21: & $\quad$$\quad$  $ \Delta^{\mathrm{backward}} = \funSym{L}^{\texSym{FS}}(\setSym{F}_{(c)}) -  \funSym{L}^{\texSym{FS}}(\setSym{F}_{(c+1)})$\\
			22: & $\quad$$\quad$ \textbf{if} ($ \Delta^{\mathrm{backward}}> v\Delta^{\mathrm{forward}}$)\\
			23: & $\quad$$\quad$ $\quad$ \textbf{break}\\
			24: & $\quad$$\quad$ \textbf{end if}\\
			25: & $\quad$ \textbf{end while}\\
			26: & $\quad$$ \setSym{F}_{(t)} = \setSym{F}_{(c+1)}  $\\
			27: & $\quad$$ t = t + 1 $\\
			28: & \textbf{end while}\\
			29: & \textbf{return} $ \setSym{F}_{(h_{i})} = \setSym{F}_{(t)}$: set of selected features\\
			\hline
		\end{tabular}
	\end{table}
	
 	\subsection{The combination of subregion and feature selections with \colorText{\acl{fbp}}}\label{subsec:ComFbp}
	
	In this part, we present the way to combine the previously proposed subregion and feature selections with two widely used \colorText{\acl{fbp}} methods, namely \acs{knn} and \acs{map}, to estimate the user location from the fingerprint $\mathbf{O}^u$ measured at the actual but unknown location $\mathbf{l}^u$. We assume that a set of candidate subregions $ \setSym{S}_m^{\mathrm{u}} $ has been selected using the \acs{mji}-based subregion selection. For each of the selected subregions $\setSym{g}_i, i\in  \setSym{S}_m^{\mathrm{u}}$, the set $ \setSym{F}_{(h^i)} $ of relevant features has been chosen using \colorRev{\acs{foba}}. While the cardinality of these sets  $ \setSym{F}_{(h^i)} $ will be different, both positioning approaches require the number of features taken into consideration to be the same for all candidate locations. So, we determine the set	
	\begin{equation}
	\label{eq:userFeatures}
	\setSym{F}_{\mathrm{u}}^{\mathrm{candidate}} = \setSymScript{A}{i}{u}\cap \setSym{F}_{\mathrm{u}}
	\end{equation}
of candidate features, where $ \setSym{F}_{\mathrm{u}} = \underset{i\in \setSym{S}_m^{\mathrm{u}}}{\cup}\setSym{F}_{(h^i)}$. The candidate features are all features actually observed by the user and available in at least one of the candidate subregions $\setSym{S}_m^{\mathrm{u}}$. We finally rank these features by the number of candidate subregions in which they are available. The $ h $ candidate  features $ \setSym{F}_{h}^{\mathrm{u}} $ ranking highest are used for \colorText{\acl{fbp}}.
	
\acs{map} uses a variety of discrete candidate locations $ \mathbf{l} $ and applies Bayes' rule to compute for each of them the degree to which the assumption that the current location of the user is $ \mathbf{l} $ is supported by the available \acs{rfm} and the currently observed fingerprint \citep{park2010growing,1498348}. Further details regarding the combination of \acs{map} with subregion and feature selection are given in \citep{Zhou2018}.

	For \acs{knn} the $k$ points $\mathbf{l}_q$ within the \acs{rfm} which are closest to the user's observation in feature space have to be identified. Assuming that their indices are collected in the set $\setSym{n}_k^u$ the estimated location $ \hat{\mathbf{l}}^{\mathrm{u}}_{k\mathrm{NN}} $ of the user is obtained from:
	\begin{equation}\label{eq:knn}
	\hat{\mathbf{l}}^{\mathrm{u}}_{k\mathrm{NN}}= \sum_{q}{\omega_{q}\mathbf{l}_q}, \forall q\in \setSym{n}_k^{\mathrm{u}},
	\end{equation}
where the respective weights $ \omega_q $ are defined \colorText{as proportional to the inverse distance of the fingerprints in the feature space:}
	\begin{equation}\label{eq:knnWeight}
	\omega_{q} = \frac{1/\funSym{d}(\setSymScript{O}{i}{u}, \setSymScript{O}{q}{})}{\sum_{p}{1/\funSym{d}(\setSymScript{O}{i}{u}, \setSymScript{O}{p}{})}}, \forall p\in \setSym{n}_k^{\mathrm{u}}.
	\end{equation}
Herein, $\funSym{d}: \setSym{O}\times\setSym{O}\mapSym\convSetSym{R}$ is a chosen distance metric, e.g., Euclidean distance or \colorText{Hamming} distance, used to measure the dissimilarity between fingerprints. The adaptation proposed by us consists in the construction of these observations $\setSym{O}$ which contain exactly one row for each of the $h$ candidate features selected previously and collected in $ \setSym{F}_{h}^{\mathrm{u}} $. Generally these vectors are much smaller than for the standard \acs{knn}-approach, where they would need to have one entry for each feature observed anywhere within the \acs{roi}. The proposed online positioning approach is summarized in \tabPref\ref{tab:onlinepositioning}.

	\begin{table}[!h]
		\centering
		\caption{Pseudocode of the proposed online positioning approach}
		\label{tab:onlinepositioning}
		\begin{tabular}{p{0.01\columnwidth}p{0.89\columnwidth}}
			\hline
			& \textbf{Algorithm}: Online positioning\\
			\hline
			1: & Input: the user's observation $\setSymScript{O}{i}{u} $;\\
			    & $\hphantom{Input: }$the selected relevant features $ \{\setSym{F}_{(h^1)},\setSym{F}_{(h^2)},\cdots,\setSym{F}_{(h^M)}\} $ of $ M $ subregions;\\
			    & $\hphantom{Input: }$the precomputed data for \acs{map} and \acs{knn}\footnote{For \acs{map}, the precomputed data are the likelihood of the selected features of each subregion and prior probability of each candidate location. For \acs{knn}, the precomputed data are the observations of the selected relevant features of each subregion.}\\
			2: & Output: the estimated location $ \hat{\mathbf{l}}^{\mathrm{u}}_i $ of the user\\
			3: & $ \setSym{S}^{\mathrm{u}}_m := $ find the $ m $ candidate subregions according to the ranking of \acs{mji}\\
			4: &  $  \setSym{F}_{h}^{\mathrm{u}}  :=$ find the candidate features according to the ranking of the frequency of the selected relevant features $ \setSym{F}_{\mathrm{candidate}}^{\mathrm{u}} $ \acs{wrt}  $ \setSym{F}_{(h^i)}, \forall i \in \setSym{S}^{\mathrm{u}}_m$\\
			5: &$\hat{\mathbf{l}}^{\mathrm{u}}_i :=$ using \acs{map} or \acs{knn} according to $ \setSym{F}_{h}^{\mathrm{u}} $ and the precomputed data\\
			6: & \textbf{return} $ \hat{\mathbf{l}}^{\mathrm{u}}_i $: the estimated location\\
			\hline
		\end{tabular}
	\end{table}

	\subsection{Computational complexity of online positioning}\label{subsec:comCom}
	
	In this part, we analyze the computational complexity of the proposed approach and compare it to \acs{map} and \acs{knn}-based positioning without subregion and feature selection. 
	
	The runtime computational complexity of estimating one position using \acs{map} and \acs{knn} are $ \mathcal{O}(\alpha M(|\setSym{A}| ^2+ 1)) $ and $ \mathcal{O}(k \alpha \cdot M\operatorname{log}(\alpha \cdot M) |\setSym{A}|) $ \footnote{\colorRev{We implemented \acs{map} as proposed in \cite{Youssef2008}. The implementation can be sped up using algebraic factorization \citep{bisio_smart_2016}.} For \acs{knn}, tree-based methods (\eg kd-tree) are used in the scikit-learn implementation \citep{scikit-learn}.}, respectively, where $ \alpha$ is the number of candidate locations in each of the $M$ subregions and $ k $ is the number of nearest neighbors used for positioning. The computational complexity of the proposed method is instead approximately equal to $ \mathcal{O}(\alpha |\setSym{S}_m^{\mathrm{u}}|\cdot|\setSym{F}_{h}^{\mathrm{u}}|^2) $ and $ \mathcal{O}(\alpha k |\setSym{S}_m^{\mathrm{u}}|\operatorname{log}(\alpha \cdot |\setSym{S}_m^{\mathrm{u}}|)\cdot|\setSym{F}_{h}^{\mathrm{u}}|) $ for \acs{map} and \acs{knn}, respectively. So, clearly the computational complexity of the proposed approach is significantly less than for the \acs{map}- and \acs{knn}-based positioning approaches without subregion and feature selection. Furthermore, it is independent of the size of the \acs{roi} and of the total number of available features within the \acs{roi}. The proposed approach is to constrain and limit the search to a set of candidate reference locations and selected features for the online positioning. Though we only give the analytical formula of the computational complexity of MAP and \acs{knn}, other fingerprinting-based location methods will also benefit from the proposed approach, because the computational complexity of \colorText{\acl{fbp}} is proportional to the size of the search space.
	
	Besides the \acs{rfm} further data required during the online positioning stage can be precomputed already during the offline stage (\figPref\ref{fig:schematic}). This holds in particular for:
	\begin{itemize}
		\item the set $\setSymScript{A}{i}{}$ of available feature keys of each subregion required for calculating the \acs{mji} at the online stage,
		\item the set $\setSym{F}_{(h^i)}$ of relevant features of each subregion calculated using \colorRev{\acs{foba}},
		\item and the conditional distribution ($ \mathrm{Prob}(\setSym{O}_j|\mathbf{l}) $) of the selected relevant features within each subregion obtained from kernel density estimation.
	\end{itemize}  
	
	At the online stage these pre-computed data are cached to the user device to achieve location estimation while realizing mobile positioning. Furthermore, only the observed values of the \colorText{features} that are selected as relevant ones in the \acs{rfm} \colorText{need to} be loaded during \colorText{the} online positioning stage. The proposed preprocessing steps also reduce the required storage space for saving the cached pre-computed data because these data only need to cover the selected relevant features instead of all the features observable within the \acs{roi}.
	
	\section{Experimental results and discussion}\label{sec:experimentalResults}
	In this section, we first describe the experimental configurations, including the \acsp{roi}, data collection and the two different types of features used, namely \acs{rss} from \acs{wlan} \acsp{ap} and \acs{ble} beacons. A detailed analysis of \acs{mji}-based subregion selection, a comparison of different feature selection methods (randomized \acs{lasso}, forward greedy, and \colorRev{\acs{foba}}), as well as the positioning accuracy and cost of time for positioning are illustrated for one \acs{roi} \colorText{using only \acs{wlan} \acsp{rss}}. Finally, we apply the proposed approach to a larger \acs{roi} \colorText{using} both types observables.
	
	\subsection{Testbed}\label{subsec:testbed}
\begin{figure}[!h]
		\centering
		\includegraphics[width=0.65\columnwidth]{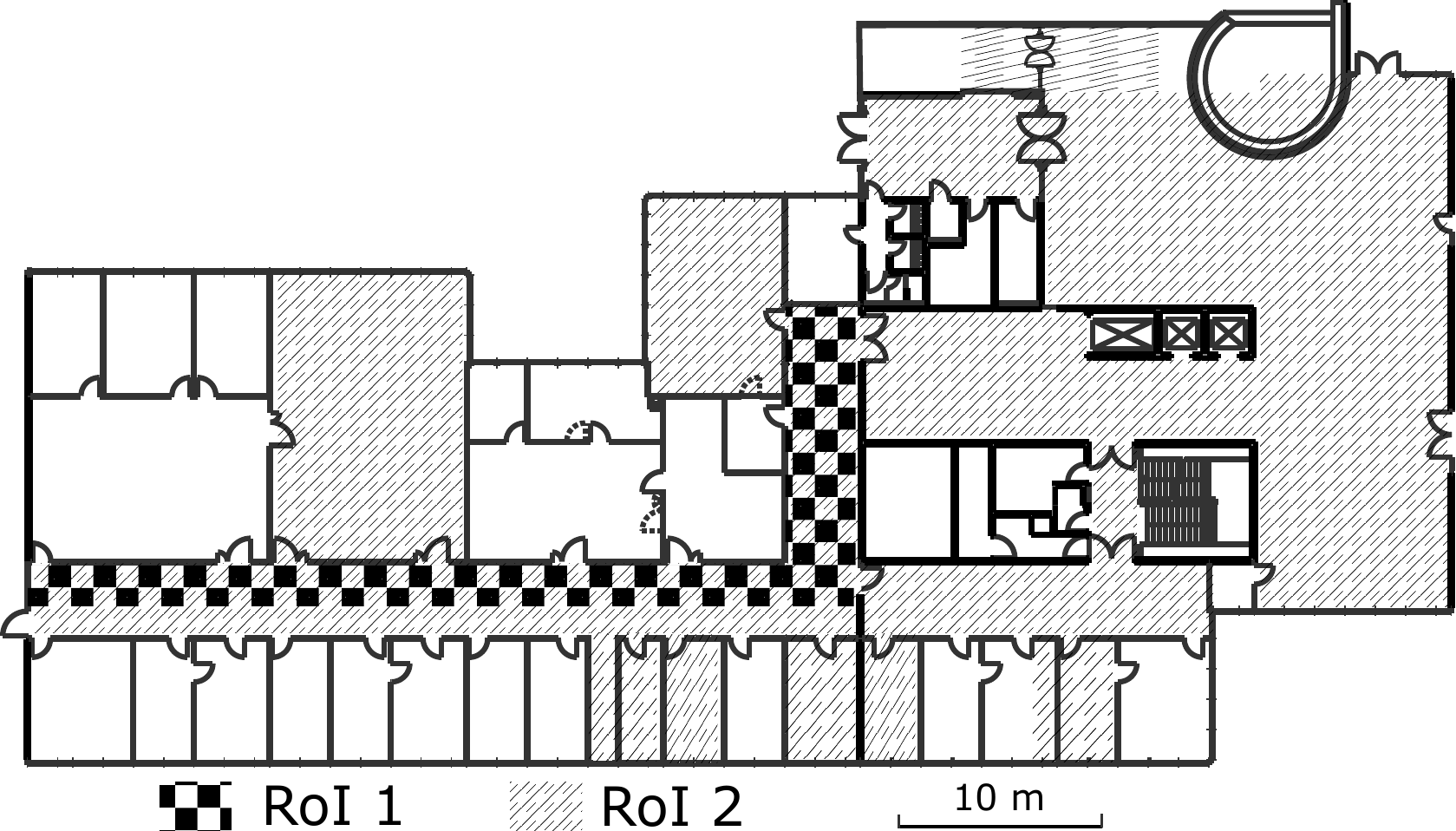}
		\caption{Two \acsp{roi} {within one floor of an office building}.}
		\label{fig:roi}
	\end{figure}

	In this section we analyze data obtained from real measurements collected using a Nexus 6P \footnote{\colorRev{Herein we carried out all experiments using only one device. It is to be expected that also the quality of the results obtained using our approach depends on the mobile device(s) used for data collection, see \citep{6595757}. However, we leave a related investigation for future work.}} smart phone (for \acs{wlan} \& \acs{ble} \acs{rss}) and a {Leica MS50} total station (for position ground truth) within two \acsp{roi} in an office building (\figPref\ref{fig:roi}) which is covered by a plethora of \acs{wlan} \acsp{ap} signals and \acs{ble} beacons\footnote{All \acsp{ap} and beacons had been deployed independently of this experiment and long before it for the purpose of internet access. Each \acs{ap} supports two frequencies (2.4 GHz and 5 GHz) and has a built-in \acs{ble} beacon. There is no automatic power adjustment of the \acsp{ap}.}.
	
	We kinematically determined the \acs{rfm} by recording data approximately every 1.5 seconds while a user walked through the \acs{roi} and a total station tracked a prism attached to the Nexus smart phone with an accuracy of about 5 mm. \colorText{The recording rate is around 0.67 Hz. This is higher than typically reported update rates for WLAN RSS and was achievable on this device by changing the hardware settings of the scanning interval. \colorRev{At the online stage, the RSS collection can be further accelerated by incremental scanning as discussed in \citep{brouwers_incremental_2014}.}} \colorText{This approach is a compromise between the high accuracy attainable by stop \& go measurements at carefully selected reference positions and} the low extra effort of crowd-sourced \acs{rfm} data collection as outlined e.g., in \citep{6817916}. \colorText{For simplicity we do not apply further pre-processing, such as filtering out the \acsp{ap} yielding low \acs{rss} values, discarding rarely measured \acsp{ap}, or merging signals transmitted from the same signal source on multi-frequencies although such steps could further reduce the computational complexity and improve the results in a real application.}
	
	In order to evaluate the performance of the proposed approach independently an additional test dataset was collected. \colorText{The coordinates of the \acfp{tp} as measured by the total station were later used as ground truth for calculating the positioning error in terms of the Euclidean distances between estimated and true coordinates. \colorRev{The use of the tracking total station for both RFM data collection and test point data collection meant that RSS data did not have to be collected at any specific points (e.g. marked ones) within the RoI or subregions, and it was not necessary to occupy the same points again.} We report the $ \texSym{50^{th}} $, $ \texSym{75^{th}} $ and $ \texSym{90^{th}} $ percentile of the horizontal positioning error (\ie \acf{ce}50, \acs{ce}75, \acs{ce}90) defined as the minimum radius for including 50\%, 75\%, and 90\% of the positioning errors \citep{Potort2018}. Furthermore, as an indicator for outlying position estimates \colorRev{(in particular due to wrong subregion selection)} we also report the percentage of positioning errors exceeding 10~m, and as an indicator of computational complexity the average time to calculate the position of the test point.} Data processing according to the proposed algorithms was implemented in Python using the scikit-learn package \citep{scikit-learn} as outlined in \figPref\ref{fig:schematic}. 
	
\begin{figure}[!htb]
		\centering
		\subfloat[Using \acs{rss} from \acs{wlan} \acsp{ap} as the features]{
			\label{subfig:roi_2_wlan_nbAvailableFeatures}
			\includegraphics[width=0.45\columnwidth]{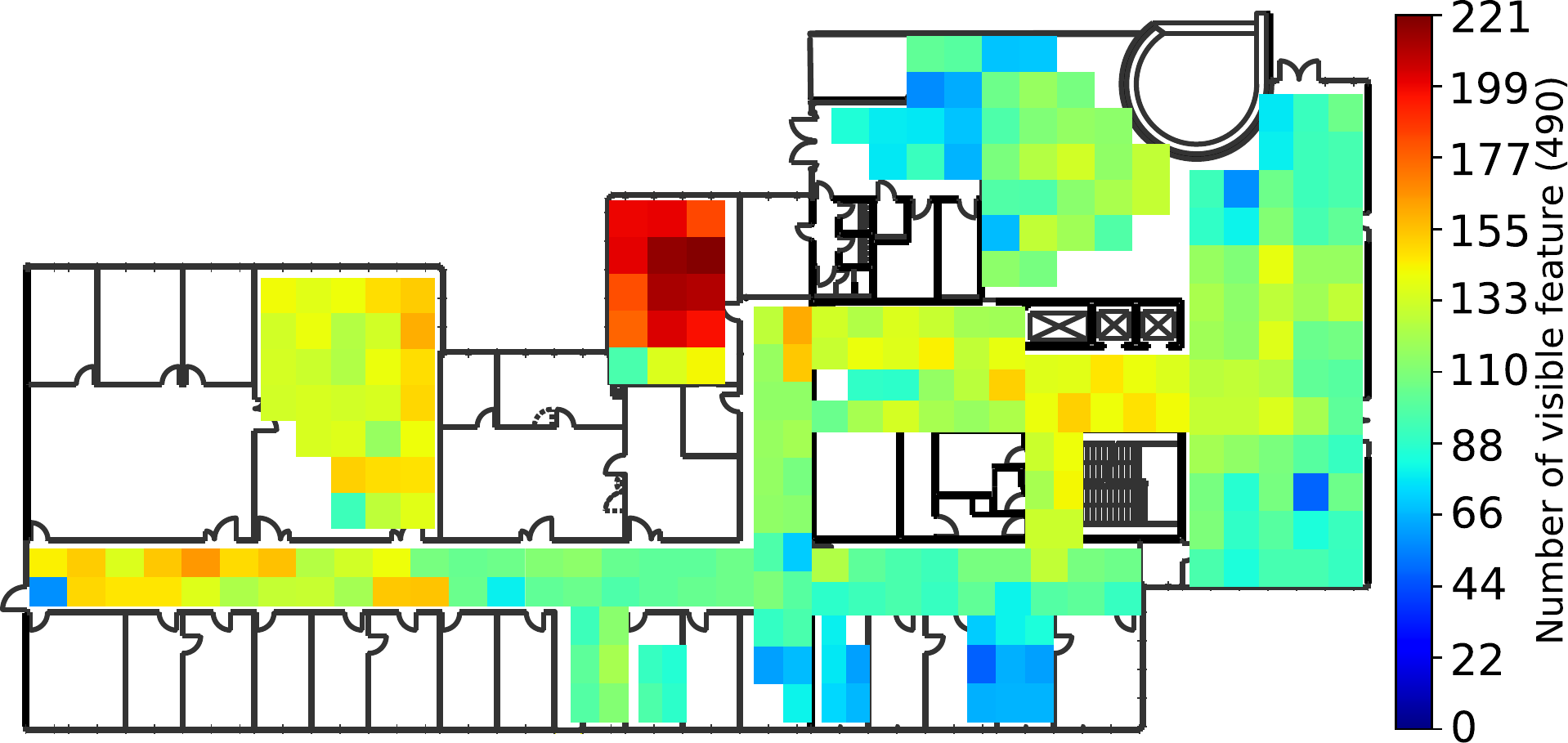}}\hspace{2ex}
		\subfloat[Using \acs{rss} from \acs{ble} beacons  as the features]{
			\label{subfig:roi_2_ble__nbAvailableFeatures}
			\includegraphics[width=0.45\columnwidth]{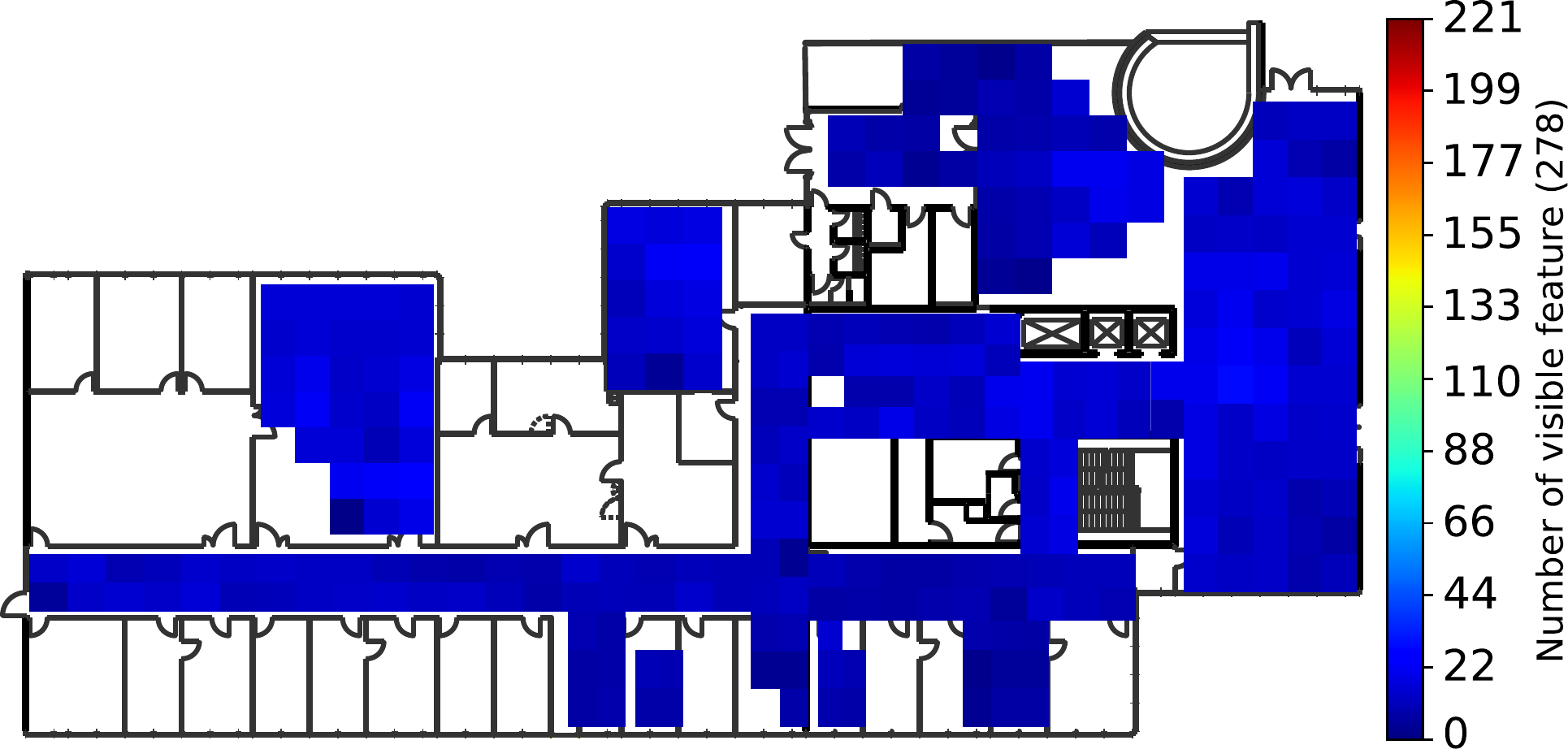}}
		\caption{The number of the available features of each subregion}
		\label{fig:roi2NbAvailableFeatures}
	\end{figure}

	The details for two \acsp{roi} are summarized in \tabPref\ref{tab:summaryRfm} and the numbers of available features (\ie visible \acs{wlan} \acsp{ap} or \acs{ble} beacons) are illustrated in \figPref\ref{fig:roi2NbAvailableFeatures}. \colorText{The number of available features changes throughout the RoI and is thus also different in the different subregions. . Abrupt changes of feature observability occur in some locations close to walls and close to support pillars or cable/pipe ducts (not contained in the available floorplan). This is caused by the uneven number of raw measurements assigned to each subregion. In \acs{roi} 1, we carried out 5-rounds of one-day data collections about one month apart. This allowed us to take both the spatial and temporal variations into account when building the \acs{rfm}.  For \acs{roi} 2, the collected \acs{rfm} is used as an example for validating the performance the proposed approach in case of a more extended \acs{roi}. Though the area of \acs{roi} 1 is a subset of \acs{roi} 2,  the two \acsp{rfm} are collected at different time and with different \acs{wlan} configurations \footnote{\colorText{An upgrade of the \acs{wlan} (\eg change \acsp{ap}) has been carried out after the data collection of \acs{roi} 1.}}.} Both \acsp{roi} are divided into subregions of size $ 2\times2 m^2$ \colorText{aligned with the floor plan of the \acs{roi}. While such an alignment may not be necessary, it is useful as it assures that individual subregions are not split by walls or other obstacles possibly causing discontinuities in the feature space. In many applications of an FIPS a floor plan exists, and can thus be used for subregion definition, because it is required for the FIPS and the associated location-based services anyway.} \colorText{In addition, some subregions contain no measurements (see \tabPref\ref{tab:summaryRfm}). We have no need to treat the empty subregions specifically because they will be filtered out by subregion selection anyway without increasing the computational burden much.}
	
\begin{figure}[!htb]
		\centering
		\colorText{
		\subfloat[Raw \acs{rss}]{
			\label{subfig:raw_rss}
			\includegraphics[width=0.45\linewidth]{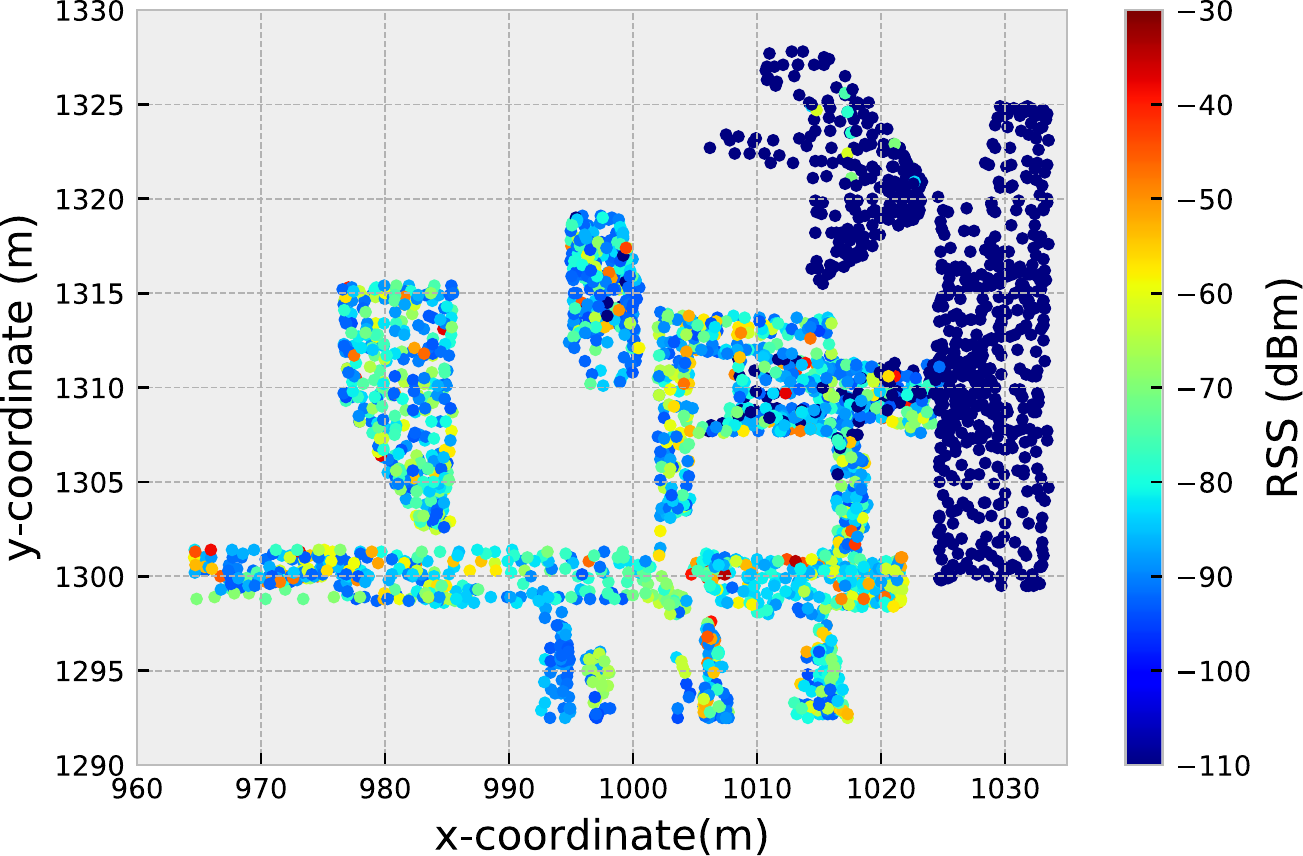}}\hspace{2ex}
		\subfloat[Kernel smoothed \acs{rss}]{
			\label{subfig:ks_rss}
			\includegraphics[width=0.45\linewidth]{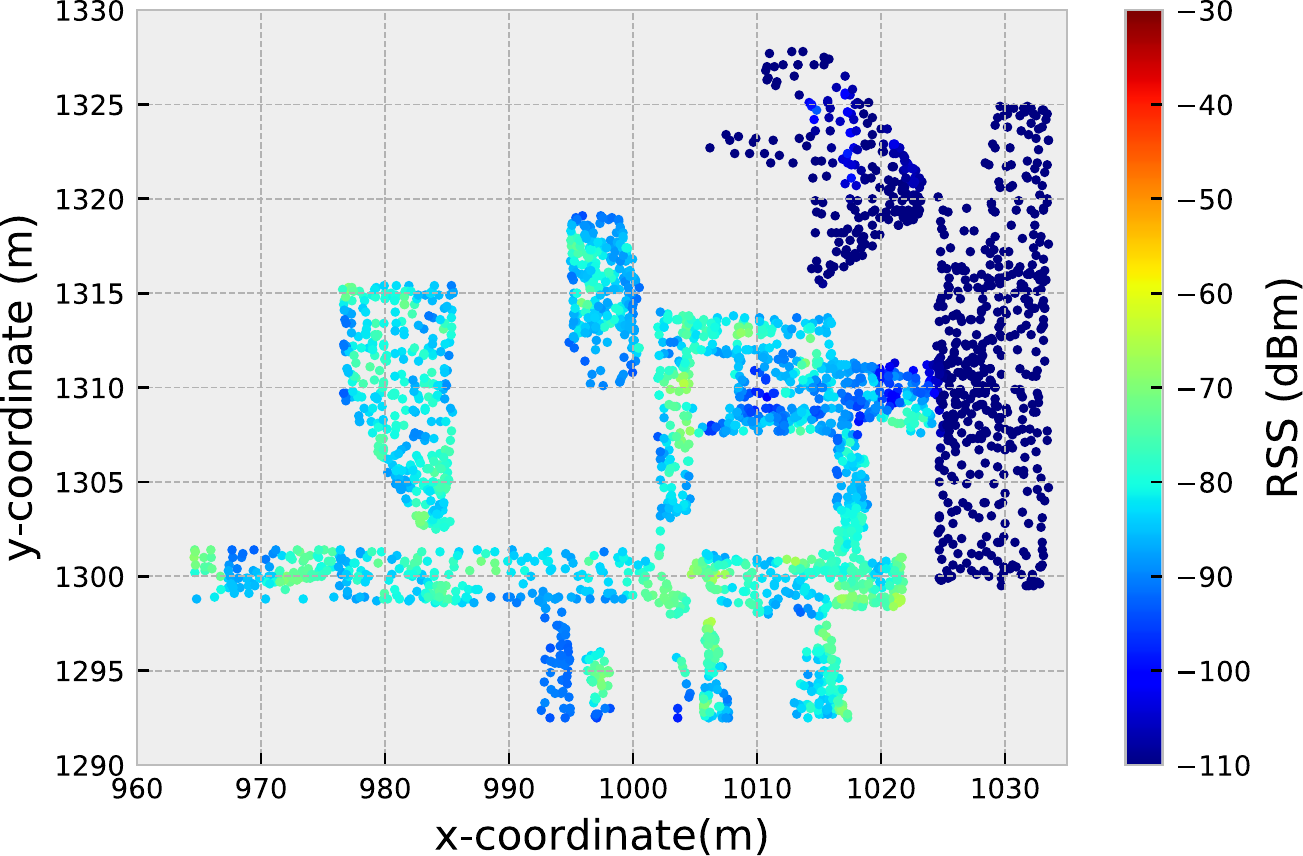}}\hspace{2ex}
		\subfloat[Interpolated \acs{rss} ( $ 0.2\times 0.2 m^2$ grid)]{
			\label{subfig:ks_interp_rss}
			\includegraphics[width=0.45\linewidth]{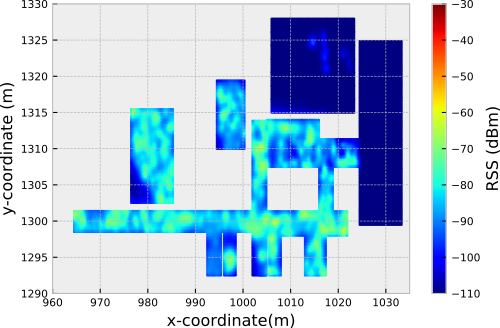}}
		\caption{An example of the spatial distribution of \acs{rss} values as used for \acs{rfm} generation}
		\label{fig:rss_raw_ks_interp}}
	\end{figure}

	The originally observed \acs{rfm} of each subregion is further densified to a regular grid of about \colorText{25} \colorText{reference} points per $ m^2 $ (\ie spacing about $ 0.2\times0.2 m^2 $) by \colorText{kernel smoothing} interpolation \colorText{(\figPref\ref{fig:rss_raw_ks_interp})} to mitigate the non-uniform point density of the original \acs{rfm} (\figPref\ref{fig:schematic}). \colorRev{This process also ensures that data are available at the same location within one subregion}. \colorText{ In this paper, we use a Mat\'ern kernel (the length scale is 1) for smoothing the spatial distribution of the \acs{rss} and assume that the propagation channel introduces the additive Gaussian white noise to the \acs{rss}.} \footnote{Assessing the impact of different uniform or non-uniform subregion shapes and sizes as well as alternative interpolation strategies is beyond the scope of this paper and left for future work.} \colorText{The smoothly interpolated raw measurements are rounded to integers for reducing the storage requirements and for mitigating the impact of the quantization of \acs{rss} values on the positioning performance \citep{Fingerprinting2017} and the \acsp{ap} are treated as non-measurable if their \acs{rss} values are lower than -100~dBm.} The resulting gridded \acs{rfm} is used for all further processing steps. 
	
	\newcolumntype{d}[1]{D{.}{.}{#1}}
	\begin{table}[!h]
		\centering
		\caption{\colorText{Summary of \acs{rfm} characteristics}}
		\label{tab:summaryRfm}
		\begin{tabular}{p{0.04\columnwidth}rp{0.15\columnwidth}p{0.1\columnwidth}p{0.1\columnwidth}p{0.1\columnwidth}p{0.08\columnwidth}}
			\hline
			\multirow{2}{*}{\acs{roi}} & \multirow{2}{*}{\shortstack{Area \\($ m^2 $)}} & \multirow{2}{*}{\shortstack{Number of\\ subregions} \footnote{Due to the constraints (\eg furnitures and decorations) of accessibility of the \acs{roi}, several subregions have no observations. The numbers in parentheses denotes the numbers of subregions containing at least one observation.}} & \multicolumn{2}{c}{Number of features} &\multirow{2}{*}{\shortstack{Training\\ data \footnote{\colorRev{The training data were obtained from the densified RFM obtained through kernel smoothing of raw RFM data, while the test data are separately collected raw data. By coincidence, the size of it is larger than that of the test data.}}}} & \multirow{2}{*}{\shortstack{Test\\ data}}\\
			& & & \acs{wlan} & \acs{ble} & & \\
			\hline
			1&120&35 (34)&399&--&1525&509\\
			2&1100&326 (285)&490&278&2855&476\\
			\hline
		\end{tabular}
	\end{table}
	
	\subsection{Analysis using \acs{wlan} signals in \acs{roi} 1}
	
	\subsubsection{Validation \acs{mji}-based subregion selection}
\begin{figure}[!h]
		\centering
		\colorText{
		\subfloat[]{
			\label{subfig:jaccard}
			\includegraphics[width=0.45\columnwidth]{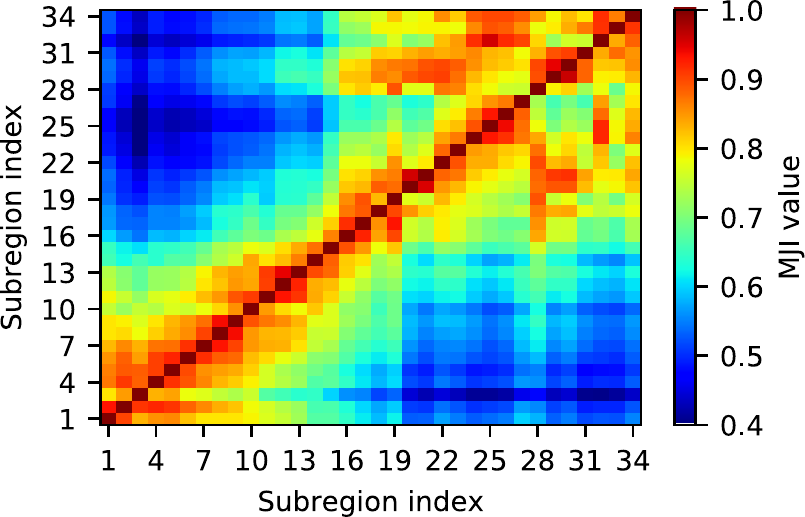}}\hspace{2ex}
		\subfloat[]{
			\label{subfig:distance}
			\includegraphics[width=0.45\columnwidth]{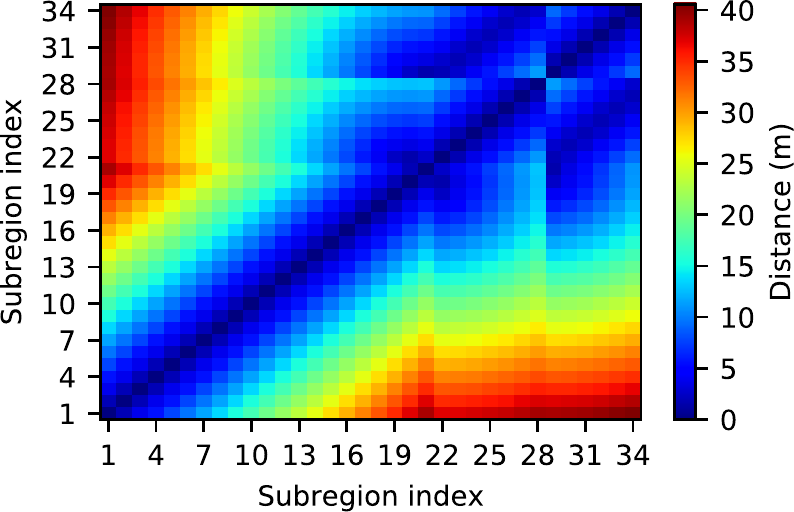}}\\
		\subfloat[]{
			\label{subfig:subregionId}
			\includegraphics[width=0.45\columnwidth]{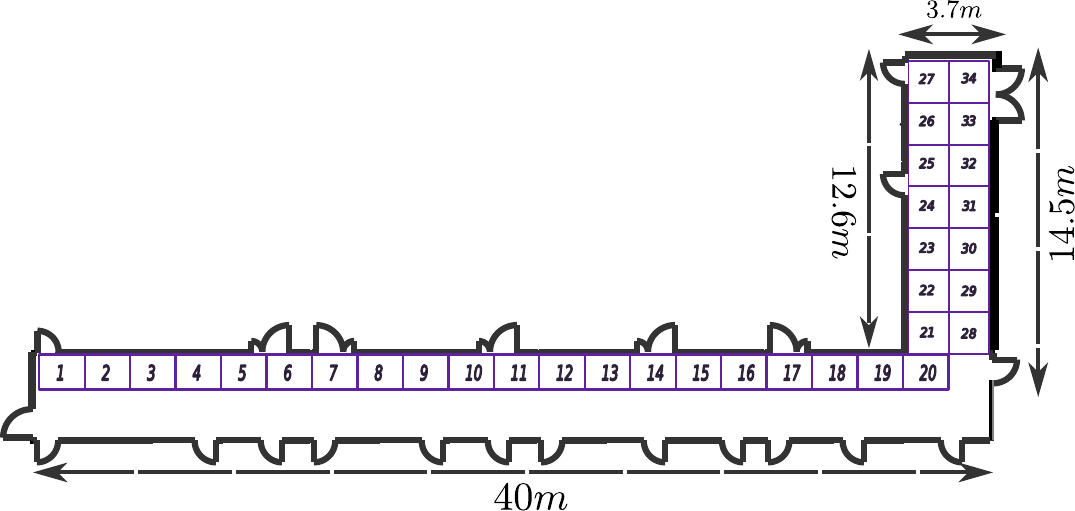}}
		\caption{Analysis of \acs{mji} between the subregions of \acs{roi} 1 (training data). (a): Spatial distribution of the \acs{mji}. \colorText{Apparent abrupt changes of \acs{mji} values appear in the up-right part. This is related to the spatial distribution of the subregions.} (b) Spatial distribution of the distance. (c) Schematic of subregion division. \colorText{Each subregion is indexed by an integer for this analysis. The concrete number associated with a specific subregion is irrelevant. }} 
		\label{fig:spatialFiltering}}
	\end{figure}

	The \acs{roi} 1 consists of 34 subregions. \figPref\ref{subfig:jaccard} shows the \acs{mji} for all pairs of subregions indicating that the index is related to the Euclidean distance of the training data. This corresponds to the expectation that the same \acsp{ap} are available in nearby subregions while different \acsp{ap} are observed in subregions far from each other.
	
	\begin{figure}[!htb]
		\centering
		\colorText{
		\subfloat[\acs{roi} 1 (\acs{wlan})]{
			\label{subfig:subregion_selection_loss_roi_1_wlan}
			\includegraphics[width=0.3\linewidth]{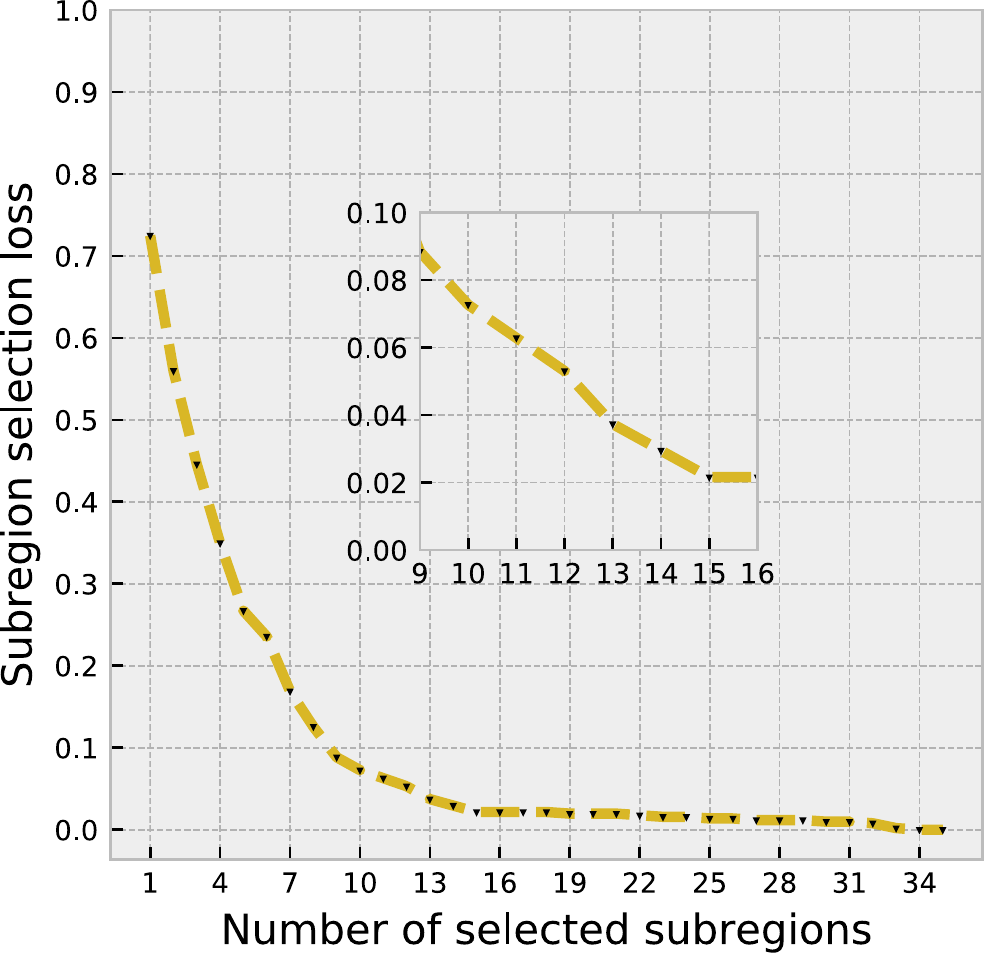}}\hspace{2ex}
		\subfloat[\acs{roi} 2  (\acs{wlan})]{
			\label{subfig:subregion_selection_loss_roi_2_wlan}
			\includegraphics[width=0.3\linewidth]{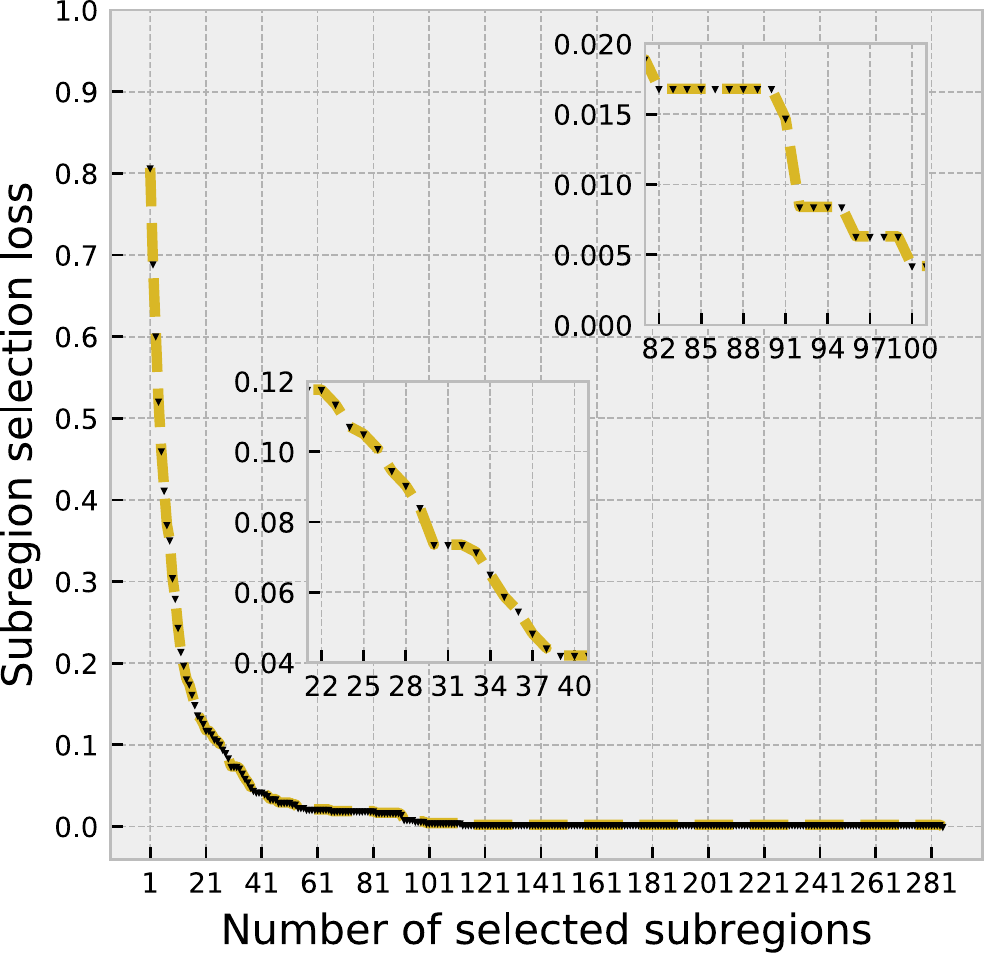}}\hspace{2ex}
		\subfloat[\acs{roi} 2  (\acs{ble})]{
			\label{subfig:subregion_selection_loss_roi_2_ble}
			\includegraphics[width=0.3\linewidth]{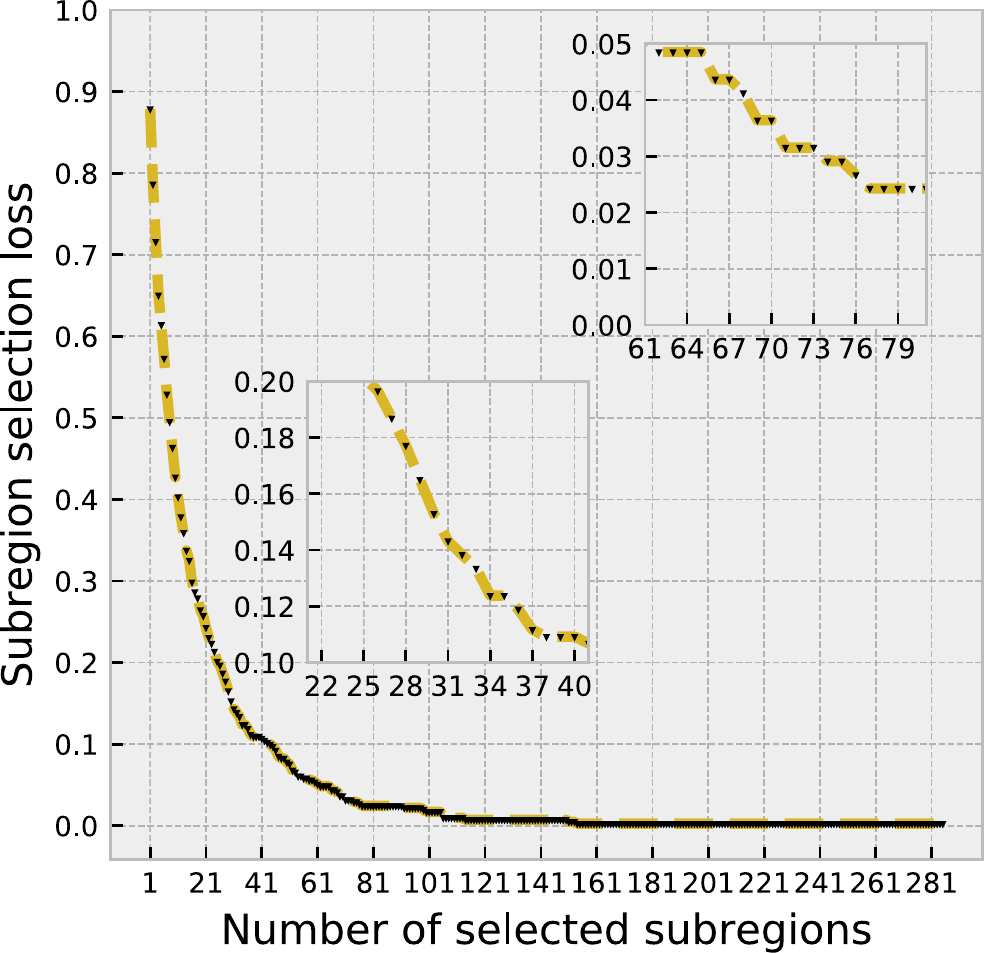}}
		\caption{The subregion selection loss with respect to the number of selected subregions}
		\label{fig:subregion_selection_losses}}
	\end{figure}
	
	The \acs{mji} value is used here as an indicator of the similarity between the features measured by the user and the ones available in the individual subregions according to the \acs{rfm}. We now use the loss function \eqref{eq:subregionSelectionLoss} to determine a suitable number $m$ of subregions for the \acs{mji}-based subregion selection. \figPref\ref{subfig:subregion_selection_loss_roi_1_wlan} shows the loss as a function of $m$ for \acs{roi} 1. The figure indicates that the loss is almost \colorText{constant if $m$ exceeds 16.}	In consecutive parts of this subsection, we analyze the performance of feature selection and positioning accuracy \acs{wrt} a given value of \colorText{$ m \in \{11, 16, 21, 34\}$}. In this range of $ m $, we show that a small compromise of subregion selection accuracy does not harm the positioning accuracy too much, which is comparable to that obtained without subregion selection, \ie for $ m =34 $ in \acs{roi} 1.
	

	\begin{savenotes}
		\begin{table}
			\centering
			\colorText{
			\caption{The selected features of subregion 17. (The unique identification of \acs{ap} is indexed from 0 to 398.)}
			\label{tab:smallRoiSelectedFeatures}
			\begin{tabular}{ccl}
				\hline
				\shortstack{Feature selection method} & \shortstack{Positioning method} & \shortstack{Selected features (top 5)}\\
				\hline
				\shortstack{Randomized \acs{lasso}} & -- &13, 107, 9, 80, 11\\ 
				\multirow{2}{*}{\shortstack{Forward greedy search}} & \acs{knn} & 11, 270, 272, 144, 19\\ 
				& \acs{map}&10, 398, 19, 20, 25 \\ 
				\multirow{2}{*}{\colorRev{\acs{foba}}} &\acs{knn}&71, 11, 270,  272, 99\\ 
				&\acs{map}&66, 71, 12, 140, 272\\ 
				\hline
			\end{tabular}}
		\end{table}
	\end{savenotes}
	
	\subsubsection{Validation of \colorRev{\acs{foba}}-based feature selection}
	
\begin{figure}[!htb]
		\centering
		\subfloat[\acs{knn}, subregion 11]{
			\label{subfig:subregion11_smallroi_knn_msepath}
			\includegraphics[width=0.3\columnwidth]{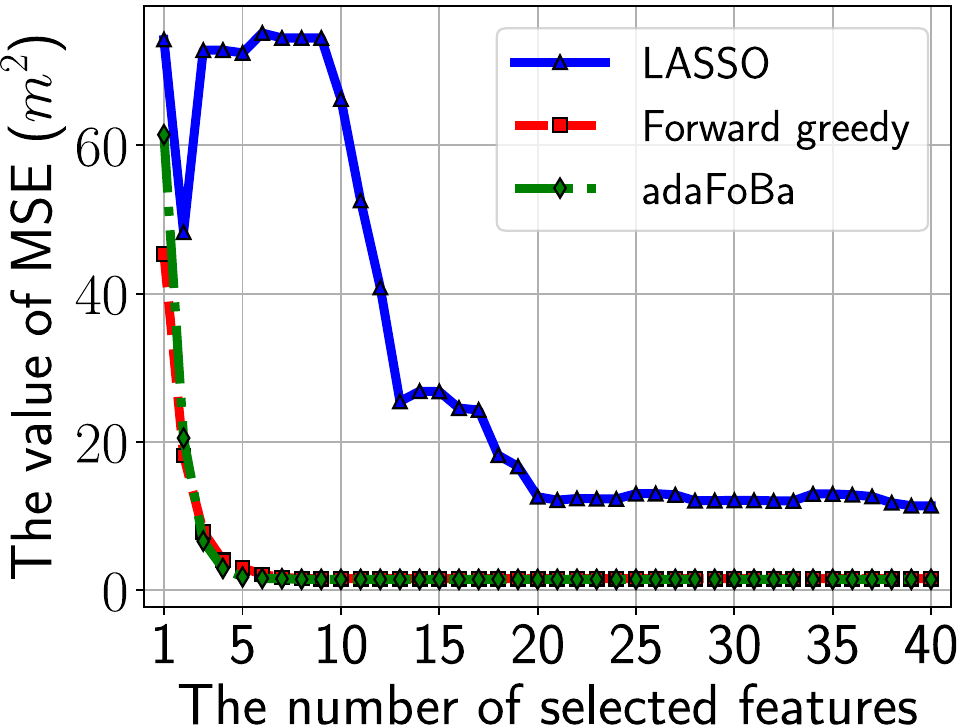}}\hspace{4ex}
		\subfloat[\acs{map}, subregion 11]{
			\label{subfig:subregion11_smallroi_map_msepath}
			\includegraphics[width=0.3\columnwidth]{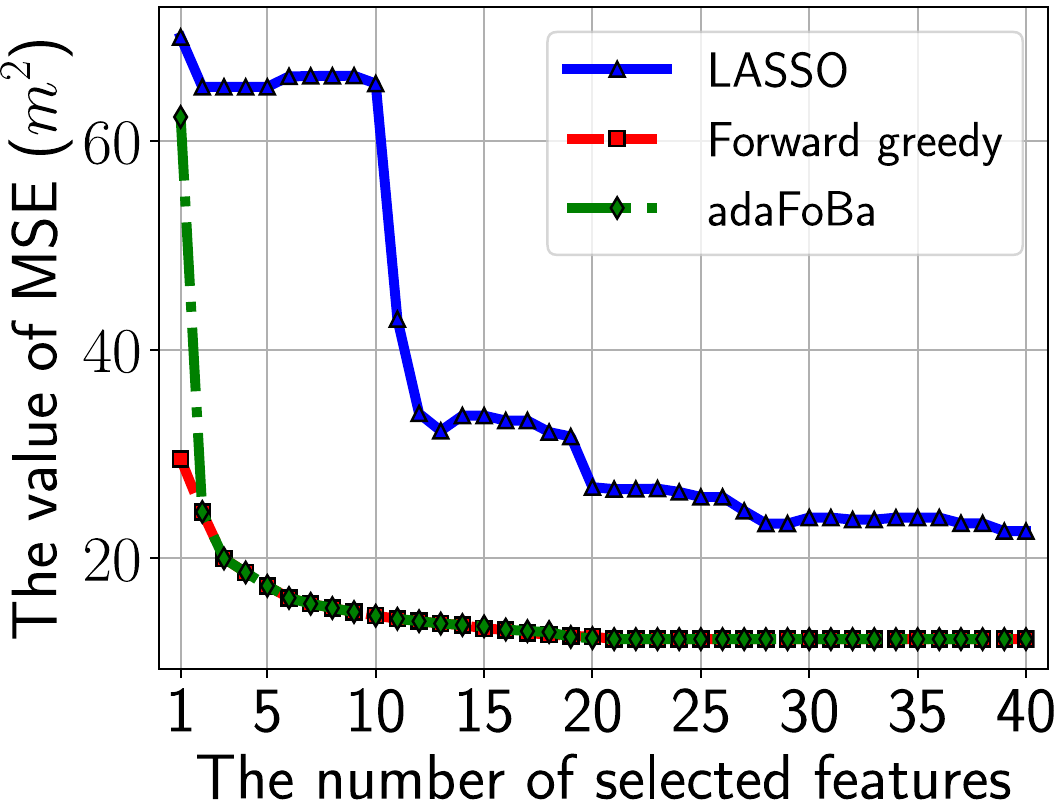}}\hspace{4ex}\\
		\subfloat[\acs{knn}, subregion 21]{
			\label{subfig:subregion21_smallroi_knn_msepath}
			\includegraphics[width=0.3\columnwidth]{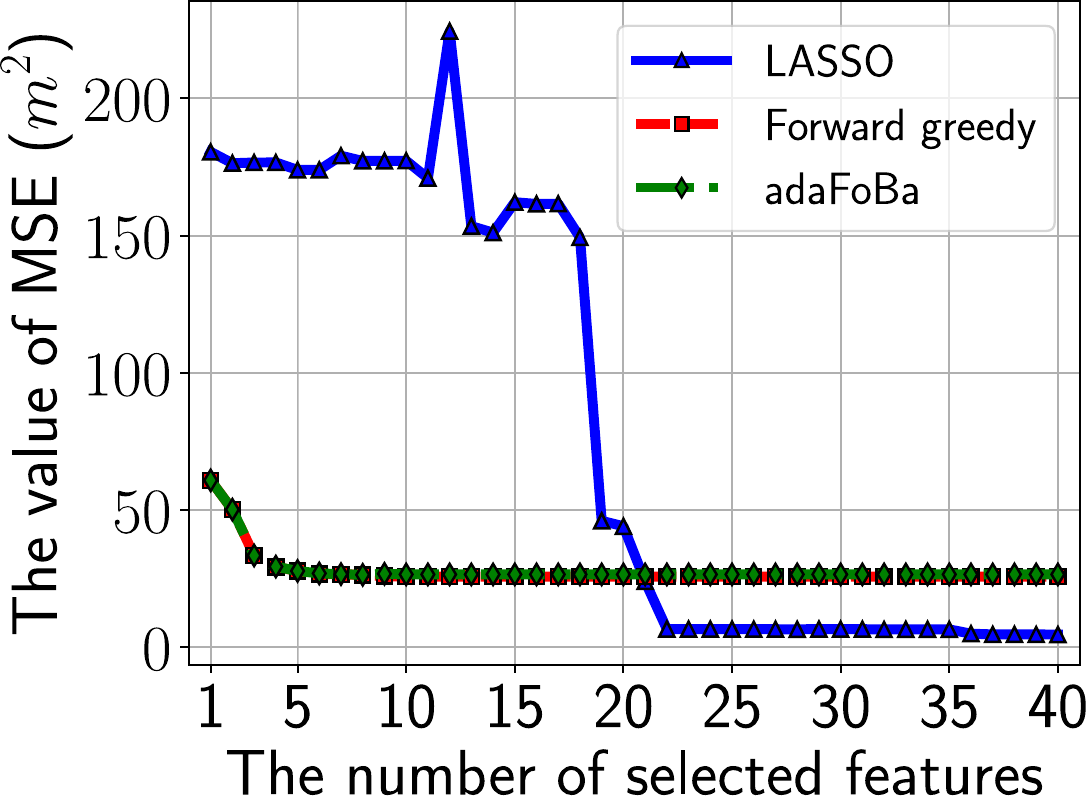}}\hspace{4ex}
		\subfloat[\acs{map}, subregion 21]{
			\label{subfig:subregion21_smallroi_map_msepath}
			\includegraphics[width=0.3\columnwidth]{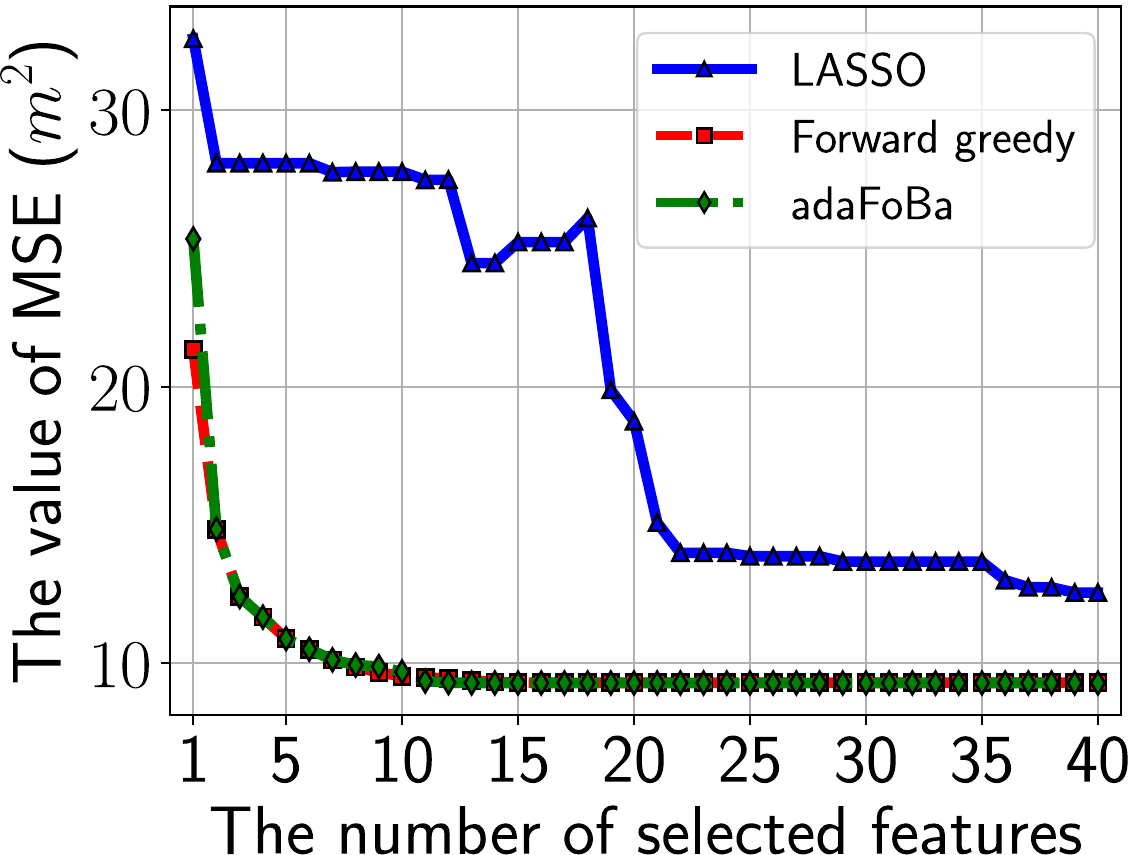}}
		\caption{The \acs{mse} paths of two subregions. In above test, the number of the selected subregion is fixed, \ie $ m = 21 $.}
		\label{fig:msePaths}
	\end{figure}

	In this part, we compare the feature selection performance using randomized \acs{lasso} as proposed in \citep{Zhou2018} to feature selection using the forward greedy algorithm and the \colorRev{\acs{foba}} proposed herein. Since the latter two are directly related to \colorText{\acl{fbp}} we evaluate the feature selection performance through the \acs{mse} of the position estimates calculated using the selected features. In \figPref\ref{fig:msePaths}, we illustrate the \acs{mse} for two arbitrarily selected subregions using \acs{knn} and \acs{map}. Regardless of the \colorText{\acl{fbp}} method and subregion, all the curves in this figure have a similar pattern. We see that i) the \acs{mse} values achieved after applying forward search and \colorRev{\acs{foba}}-based feature selection converge faster and more consistently than after applying randomized \acs{lasso}-based feature selection, and ii) randomized \acs{lasso}-based selection performance stabilizes with only a large number of selected features (\eg $ > 20 $). One explanation of this pattern is that randomized \acs{lasso} selects the features based on a regularized linear regression model instead of taking the \colorText{\acl{fbp}} methods into account. However, if a higher number of features is used for positioning, the contribution of feature selection becomes less critical.
	
	The difference of \acs{mse} between forward greedy search and \colorRev{\acs{foba}}-based feature selection is very small (\figPref\ref{fig:msePaths}), because the selected features are very similar in both cases, see \tabPref\ref{tab:smallRoiSelectedFeatures}.  However, the \colorRev{\acs{foba}}-based results are slightly better and we thus recommend it because of the increased flexibility.
	\subsubsection{The processing time and positioning accuracy}
	\tabPref\ref{tab:cmpProcessTime_CPA} presents the processing time\footnote{We used Python to implement the proposed method and evaluate the processing time using the \textit{time} package (https://docs.python.org/3/library/time.html\#module-time).} for positioning one \acs{tp} of \acsp{roi} 1 using \acs{wlan} signals as the features. \colorText{The positioning time of \acs{map} based on the \colorRev{\acs{foba}} selected features within {21 selected subregions} is about 2.9 seconds, which is almost 10 times} faster than that of using all features for positioning by searching over all the subregions. The positioning time is also lower than when using \acs{lasso} for feature selection. One explanation is that a lower number of features is selected as relevant by the proposed algorithms than by \acs{lasso}, as shown in \figPref\ref{fig:msePaths}. \colorText{As for the positioning performance, the $ \texSym{90^{th}} $ percentile of the circular error (CE90) increases by less than 0.7~m for both \acs{map} and \acs{map} based on the \colorRev{\acs{foba}} selected features within 9 selected subregions as compared to that of using all features for positioning and searching over the whole \acs{roi}. In addition, the resampling and the subregion selection reduce the percentage of large errors. So the reduced processing effort comes at the price of a small loss in accuracy.} \colorRev{If need be, the percentage of outlying position estimates could be further reduced by position filtering taking the user's motion or prior knowledge like floor plans into account during subregion selection and position estimation. This is beyond the scope of this paper and thus not further investigated here.}
	

	\begin{table*}
		\centering
		\colorText{
			\scriptsize
			\caption{{Comparison of positioning times and accuracies of \acs{roi} 1} ($ h=-1 $ denotes that the all the relevant features within the selected subregions are used for positioning. These features are selected adaptively by \colorRev{\acs{foba}} (see \secPref\ref{subsec:adaFoba}) ; their number is thus different for different subregions.) }
			\label{tab:cmpProcessTime_CPA}
			\begin{tabular}{clrcccc}
				\hline\noalign{\smallskip}
				{Methods}&{\shortstack{Values of\\$ (m, h) $}} & {\shortstack{Mean positioning \\time (\textit{s})}} & {\acs{ce}50 (m)}&{\acs{ce}75 (m)}& {\acs{ce}90 (m)} & {\shortstack{Ratio of\\ large errors} (\%)}\\
				\noalign{\smallskip}\hline\noalign{\smallskip}
				\acs{map} (without interpolation)
				&all &1.6&2.4&5.1&8.0&6.9\\
				\acs{knn} (without interpolation)
				&all &$ 2.1 \times 10^{-2} $&1.8&3.8&7.0&4.5\\
				\noalign{\smallskip}\hline\noalign{\smallskip}
				\acs{map} (with interpolation)
				&all &34.7&2.2&4.9&8.0&5.5\\
				\acs{knn} (with interpolation)
				&all &0.1&1.8&4.0&6.8&3.7\\
				\noalign{\smallskip}\hline\noalign{\smallskip}
				\multirow{4}{*}{\acs{map} (\acs{lasso})} 
				&(11, -1)&2.4&2.9&5.4&9.0&8.6\\
				&(16, -1)&4.3&2.6&5.3&9.2&8.6\\
				&(21, -1)&6.5&2.7&5.3&9.2&8.6\\
				&(34, -1)&11.6&2.5&5.0&8.8&8.6\\
				&(34, 399)&34.7&2.2&4.9&8.0&5.5\\
				\noalign{\smallskip}\hline\noalign{\smallskip}
				\multirow{4}{*}{\shortstack{\acs{knn}} (\acs{lasso})} 
				&(11, -1)&0.3$\times 10^{-2}$&2.2&4.5&7.5&4.1\\
				&(16, -1)&0.5$\times 10^{-2}$&2.3&4.3&7.4&4.7\\
				&(21, -1)&0.6$\times 10^{-2}$&2.0&4.2&7.3&4.9\\
				&(34, -1)&1.0$\times 10^{-2}$&2.2&4.0&7.1&4.7\\
				&(34, 399)&1.0$\times 10^{-2}$&1.8&3.9&6.9&4.5\\
				\noalign{\smallskip}\hline\noalign{\smallskip}
				\multirow{4}{*}{\shortstack{\acs{map} \\(Forward\\greedy search)}} 
				&(11, -1)&1.4&2.7&5.3&8.6&7.1\\
				&(16, -1)&2.5&2.7&5.6&9.2&8.4\\
				&(21, -1)&3.9&2.7&5.6&9.4&8.8\\
				&(34, -1)&6.9&2.7&5.5&9.3&8.3\\
				&(34, 399)&34.9&2.4&5.2&8.2&6.3\\
				\noalign{\smallskip}\hline\noalign{\smallskip}
				\multirow{4}{*}{\shortstack{\acs{knn} \\ (Forward\\greedy search)} } 
				&(11, -1)&0.2$\times 10^{-2}$&2.5&4.8&7.5&4.7\\
				&(16, -1)&0.3$\times 10^{-2}$&2.8&5.1&8.2&5.5\\
				&(21, -1)&0.3$\times 10^{-2}$&2.5&4.7&7.7&5.7\\
				&(34, -1)&0.5$\times 10^{-2}$&2.3&4.7&8.4&5.3\\
				&(34, 399)&1.0$\times 10^{-2}$&1.9&4.0&7.1&4.1\\
				\noalign{\smallskip}\hline\noalign{\smallskip}
				\multirow{4}{*}{\shortstack{\acs{map} \\(\colorRev{\acs{foba}})}} 
				&(11, -1)&1.3&2.6&5.0&8.6&6.9\\
				&(16, -1)&2.4&2.7&5.5&9.2&8.1\\
				&(21, -1)&3.7&2.8&5.5&9.1&8.4\\
				&(34, -1)&6.6&2.7&5.7&9.2&8.3\\
				&(34, 399)&35.0&2.3&5.2&8.0&5.9\\
				\noalign{\smallskip}\hline\noalign{\smallskip}
				\multirow{4}{*}{\shortstack{\acs{knn} \\ (\colorRev{\acs{foba}})} } 
				&(11, -1)&0.2$\times 10^{-2}$&2.9&5.0&7.4&4.5\\
				&(16, -1)&0.3$\times 10^{-2}$&3.2&5.5&8.2&4.5\\
				&(21, -1)&0.3$\times 10^{-2}$&2.7&5.3&8.1&5.1\\
				&(34, -1)&0.4$\times 10^{-2}$&2.8&5.0&8.7&7.1\\
				&(34, 399)&1.0$\times 10^{-2}$&1.8&4.0&6.8&3.7\\
				\noalign{\smallskip}\hline\noalign{\smallskip}
		\end{tabular}}
	\end{table*}

	\subsection{Analysis using WLAN and BLE signals in \acs{roi} 2}
	
	In the larger \acs{roi} 2 both \acs{wlan} and \acs{ble} signals are extracted as fingerprints. In this subsection, we present the results of applying the proposed algorithm to that \acs{roi} and both signal types. \colorText{As shown in \figPref\ref{subfig:subregion_selection_loss_roi_2_wlan} and \figPref\ref{subfig:subregion_selection_loss_roi_2_ble}, \acs{mji}-based subregion selection is applicable also in this case. The figure indicates that there is no need to search within all subregions, but actually a subset is sufficient. Appropriate numbers of the selected subregions are 32 and 38 for using \acs{wlan} and \acs{ble} {signals, respectively.}} These search spaces are less than 12\% of the area of the \acs{roi}.
	
	\begin{figure}[!htb]
		\centering
		\colorText{
			\subfloat[\acs{wlan}]{
			\label{subfig:roi_2_wlan_subregion_selection_large_error_example}
			\includegraphics[width=0.4\columnwidth]{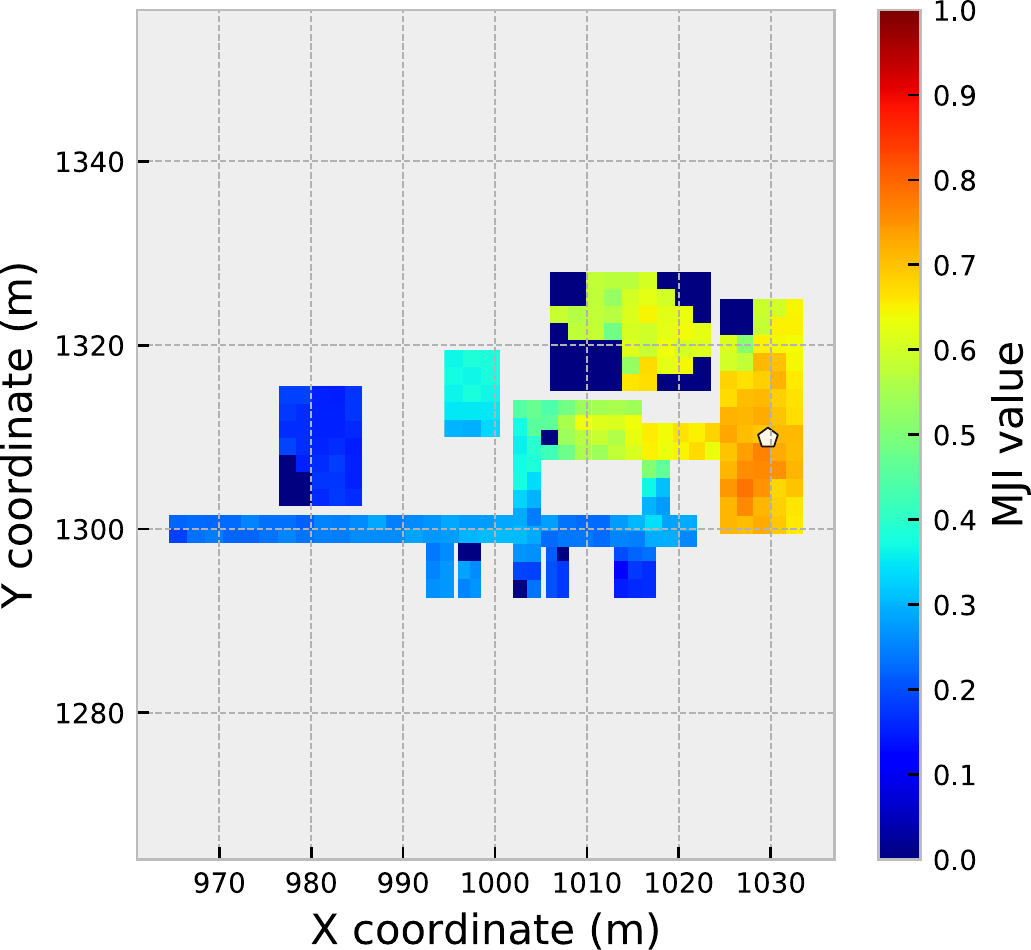}}\hspace{4ex}
		\subfloat[\acs{ble}]{
			\label{subfig:roi_2_ble_subregion_selection_large_error_example}
			\includegraphics[width=0.4\columnwidth]{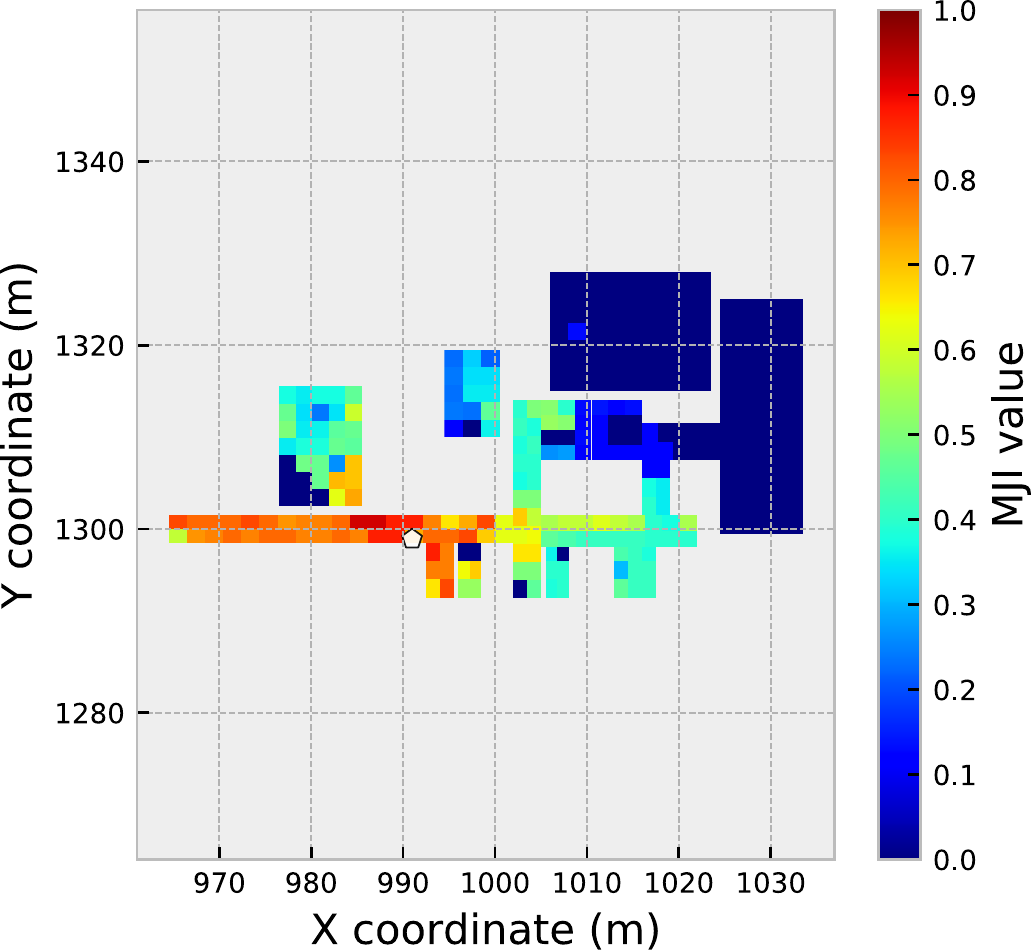}}
		\caption{Examples resulting in large subregion selection error. In the figure, the white color filled pentagon indicates the ground truth of the \acsp{tp}.}
		\label{fig:roi2SubregionSelection_large_error_Examples}}
	\end{figure}

	\begin{figure}[!htb]
	\centering
	\colorText{
		\subfloat[\acs{wlan}]{
			\label{subfig:roi_2_wlan_subregion_selection_small_errror_example}
			\includegraphics[width=0.4\columnwidth]{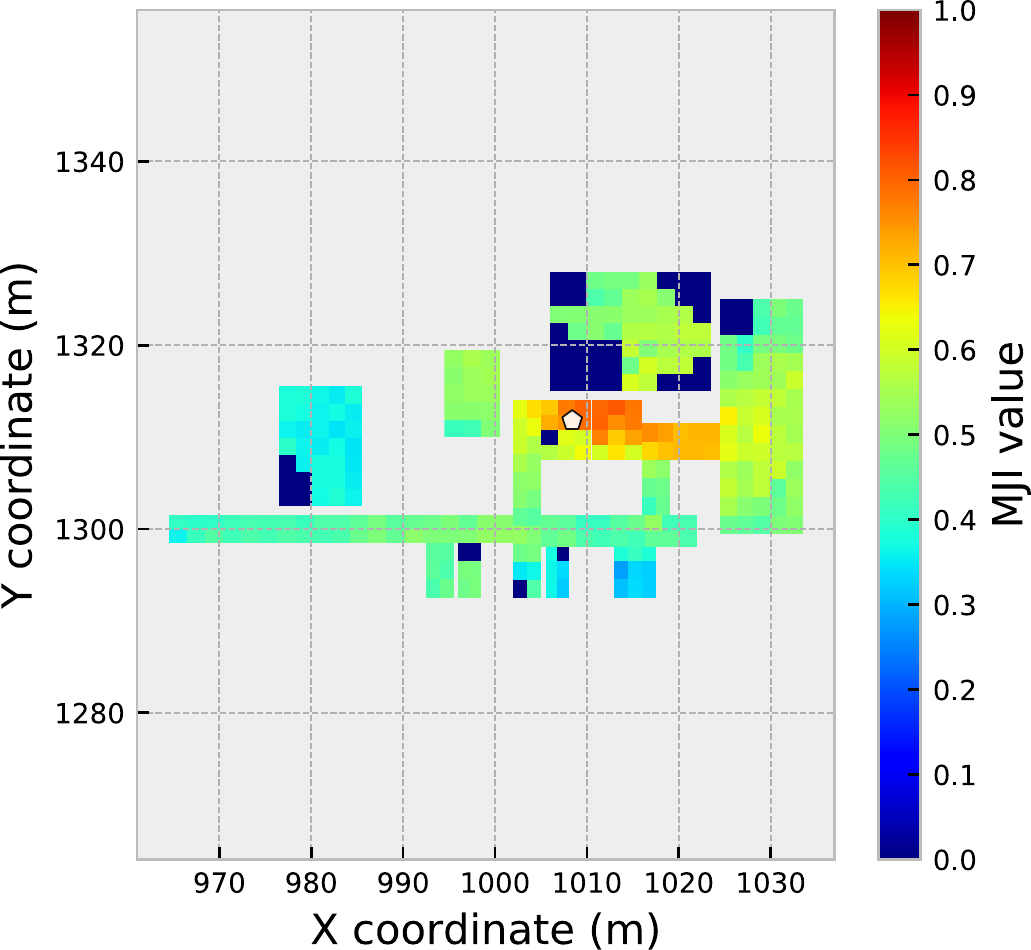}}\hspace{4ex}
		\subfloat[\acs{ble}]{
			\label{subfig:roi_2_ble_subregion_selection_small_errror_example}
			\includegraphics[width=0.4\columnwidth]{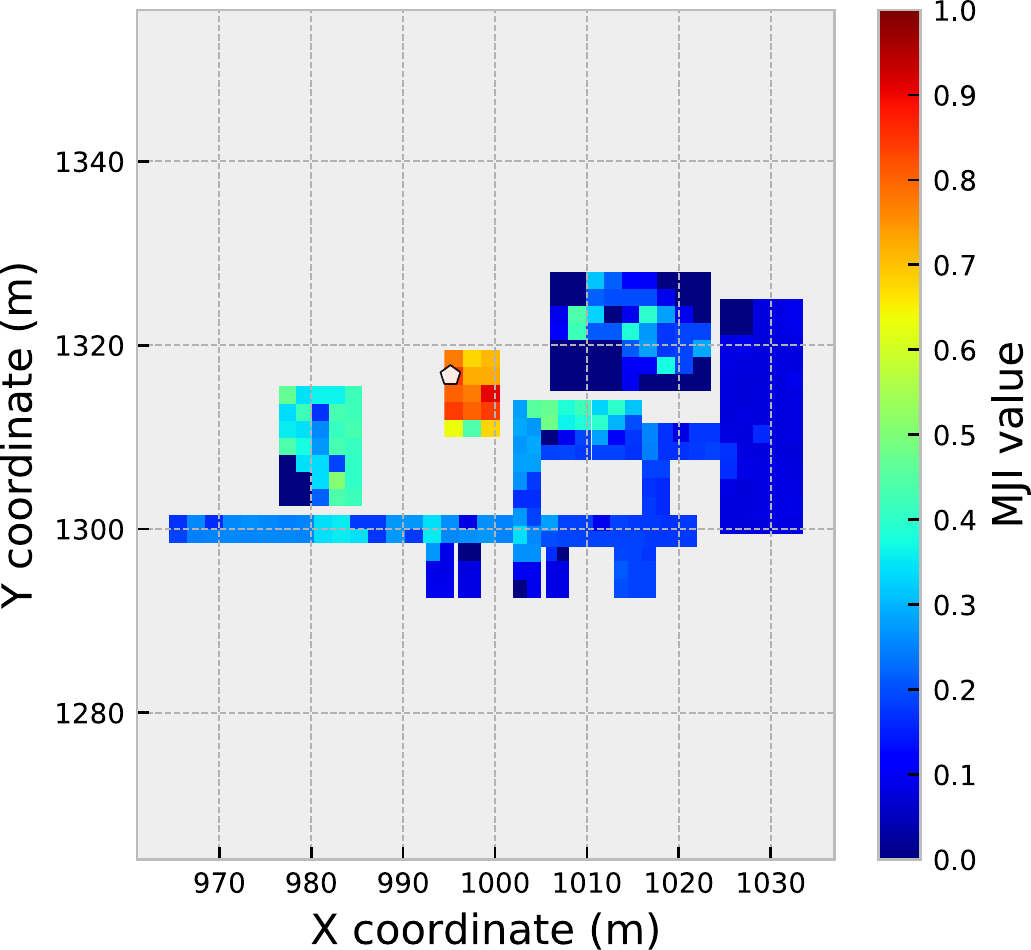}}
		\caption{Examples resulting in small subregion selection error. In the figure, the white color filled pentagon indicates the ground truth of the \acsp{tp}.}
		\label{fig:roi2SubregionSelection_small_errror_Examples}}
	\end{figure}
	
	\colorText{\figPref\ref{fig:subregion_selection_losses} also shows that there is a small fraction of test points (e.g. about 7\% for WLAN signals and RoI 2) for which the correct subregion is not among the selected ones even when choosing 30 subregions. This may seem astonishing, but an investigation of the related test cases (see e.g., \figPref\ref{fig:roi2SubregionSelection_small_errror_Examples}) shows that this occurs mostly in cases where the gradient of the MJI within an extended neighborhood of the test point is small. In that case a few or individual additional or missing features can significantly influence the ranking of the subregions in terms of MJI. We observe and need to expect this case in larger, unobstructed areas like a hall without large furniture where there is little variability in the IDs of the features observed (e.g., the visible APs in case of WLAN RSS). This characteristic suggests that an augmented scheme of selecting the subregions is required for future work, e.g. selecting an adaptive number of subregions according to the spatial gradient of the MJI, dividing the subregions of varying size, or including the feature values for subregion selection in areas with little variability.}
	
	
	\colorText{We present the positioning results of \acs{map} and \acs{knn} using \acs{wlan} and \acs{ble} signals as the features in \tabPref\ref{tab:cmpProcessTime_CPA_roi_2_wlan} and \tabPref\ref{tab:cmpProcessTime_CPA_roi_2_ble}, respectively . We conclude from the results that the proposed subregion and feature selections are beneficial for the positioning with respect to constraining the online positioning by reducing the processing time and increasing the positioning accuracy. In fact, the circular errors CE50, CE75, and CE90 as defined above get smaller and the percentage of errors larger than 10~m reduces. } 
	
	
	\begin{table*}
		\centering
		\colorText{
			\scriptsize
			\caption{{Comparison of positioning times and accuracies of \acs{roi} 2 using \acs{wlan} signal as the features} ($ h=-1 $ has the same meaning as in \tabPref\ref{tab:cmpProcessTime_CPA}.) }
			\label{tab:cmpProcessTime_CPA_roi_2_wlan}
			\begin{tabular}{clccccc}
				\hline\noalign{\smallskip}
				{Methods}&{\shortstack{Values of\\$ (m, h) $}} & {\shortstack{Mean positioning \\time (\textit{s})}} & {\acs{ce}50 (m)}&{\acs{ce}75 (m)}& {\acs{ce}90 (m)} & {\shortstack{Ratio of\\ large errors} (\%)}\\
				\noalign{\smallskip}\hline\noalign{\smallskip}
				\acs{map} (without interpolation)
				& all &17.9&3.3&6.5&9.7&9.3\\
				\acs{knn} (without interpolation)
				& all &0.1&1.6&3.1&4.8&1.7\\
				\noalign{\smallskip}\hline\noalign{\smallskip}
				\acs{map} (with interpolation)
				&all &337.5&2.6&5.6&8.6&7.4\\
				\acs{knn} (with interpolation)
				&all &0.30&1.4&2.6&4.6&1.9\\
				\noalign{\smallskip}\hline\noalign{\smallskip}
				\multirow{4}{*}{\acs{map} (\acs{lasso})} 
				&(11, -1)&3.8&3.1&5.6&8.7&7.4\\
				&(16, -1)&7.0&3.3&5.9&8.7&6.5\\
				&(21, -1)&10.0&3.4&6.4&9.2&8.0\\
				&(34, -1)&20.3&3.2&6.4&9.4&8.6\\
				&(34, 490)&43.7&2.6&5.1&8.1&6.1\\
				\noalign{\smallskip}\hline\noalign{\smallskip}
				\multirow{4}{*}{\shortstack{\acs{knn}} (\acs{lasso})} 
				&(11, -1)&0.6$\times 10^{-2}$&2.3&4.2&7.2&4.2\\
				&(16, -1)&0.8$\times 10^{-2}$&2.3&4.3&6.8&4.0\\
				&(21, -1)&1.0$\times 10^{-2}$&2.2&4.1&6.9&4.6\\
				&(34, -1)&1.6$\times 10^{-2}$&2.3&4.2&6.6&4.4\\
				&(34, 490)&1.3$\times 10^{-2}$&1.3&2.6&4.7&1.9\\
				\noalign{\smallskip}\hline\noalign{\smallskip}
				\multirow{4}{*}{\shortstack{\acs{map} \\(Forward\\greedy search)}} 
				&(11, -1)&3.5&2.4&4.4&7.5&4.6\\
				&(16, -1)&6.2&2.4&4.7&7.7&5.0\\
				&(21, -1)&9.2&2.5&4.7&7.6&4.4\\
				&(34, -1)&17.7&2.7&5.4&8.6&6.5\\
				&(34, 490)&43.0&2.7&5.1&8.1&6.1\\
				\noalign{\smallskip}\hline\noalign{\smallskip}
				\multirow{4}{*}{\shortstack{\acs{knn} \\ (Forward\\greedy search)} } 
				&(11, -1)&0.4$\times 10^{-2}$&2.5&4.3&7.0&3.4\\
				&(16, -1)&0.5$\times 10^{-2}$&2.3&4.5&6.9&5.5\\
				&(21, -1)&0.6$\times 10^{-2}$&2.2&4.3&7.4&5.7\\
				&(34, -1)&0.8$\times 10^{-2}$&2.0&3.9&6.3&3.8\\
				&(34, 490)&1.3$\times 10^{-2}$&1.3&2.6&4.3&2.1\\
				\noalign{\smallskip}\hline\noalign{\smallskip}
				\multirow{4}{*}{\shortstack{\acs{map} \\(\colorRev{\acs{foba}})}} 
				&(11, -1)&3.4&2.6&4.6&7.4&5.0\\
				&(16, -1)&6.4&2.6&5.2&7.9&5.9\\
				&(21, -1)&10.3&2.7&5.1&7.9&5.7\\
				&(34, -1)&22.5&2.8&5.2&8.4&5.5\\
				&(34, 490)&46.2&2.7&5.0&8.0&5.9\\
				\noalign{\smallskip}\hline\noalign{\smallskip}
				\multirow{4}{*}{\shortstack{\acs{knn} \\ (\colorRev{\acs{foba}})} } 
				&(11, -1)&0.6$\times 10^{-2}$&2.5&4.1&6.8&3.6\\
				&(16, -1)&0.7$\times 10^{-2}$&2.5&4.5&7.2&4.2\\
				&(21, -1)&0.8$\times 10^{-2}$&2.4&4.3&6.8&3.6\\
				&(34, -1)&1.2$\times 10^{-2}$&2.0&4.0&6.4&2.5\\
				&(34, 490)&2.3$\times 10^{-2}$&1.3&2.6&4.5&2.1\\
				\noalign{\smallskip}\hline\noalign{\smallskip}
		\end{tabular}}
	\end{table*}

	\begin{table*}
		\centering
		\colorText{
			\scriptsize
			\caption{{Comparison of positioning times and accuracies of \acs{roi} 2 using \acs{ble} signal as the features} ($ h=-1 $ has the same meaning as in \tabPref\ref{tab:cmpProcessTime_CPA}.) }
			\label{tab:cmpProcessTime_CPA_roi_2_ble}
			\begin{tabular}{clccccc}
				\hline\noalign{\smallskip}
				{Methods}&{\shortstack{Values of\\$ (m, h) $}} & {\shortstack{Mean positioning \\time (\textit{s})}} & {\acs{ce}50 (m)}&{\acs{ce}75 (m)}& {\acs{ce}90 (m)} & {\shortstack{Ratio of\\ large errors} (\%)}\\
				\noalign{\smallskip}\hline\noalign{\smallskip}
				\acs{map} (without interpolation)
				&all &5.9&5.2&9.7&17.1&23.6\\
				\acs{knn} (without interpolation)
				&all &$ 2.6 \times 10^{-2} $&3.4&6.5&9.8&9.5\\
				\noalign{\smallskip}\hline\noalign{\smallskip}
				\acs{map} (with interpolation)
				&all &197.0&4.7&8.6&12.3&13.7\\
				\acs{knn} (with interpolation)
				&all &0.1&3.4&6.9&11.3&12.2\\
				\noalign{\smallskip}\hline\noalign{\smallskip}
				\multirow{4}{*}{\acs{map} (\acs{lasso})} 
				&(11, -1)&2.0&3.8&7.8&11.0&13.1\\
				&(16, -1)&3.1&4.0&8.0&11.3&12.9\\
				&(21, -1)&4.4&4.2&8.0&11.7&14.4\\
				&(34, -1)&8.0&4.2&8.1&12.5&15.3\\
				&(34, 278)&23.5&4.3&8.4&13.4&15.8\\
				\noalign{\smallskip}\hline\noalign{\smallskip}
				\multirow{4}{*}{\shortstack{\acs{knn}} (\acs{lasso})} 
				&(11, -1)&0.4$\times 10^{-2}$&3.3&6.0&10.2&10.9\\
				&(16, -1)&0.5$\times 10^{-2}$&3.2&6.6&10.3&11.2\\
				&(21, -1)&0.6$\times 10^{-2}$&3.3&6.3&10.4&11.2\\
				&(34, -1)&0.9$\times 10^{-2}$&3.3&6.5&10.5&11.4\\
				&(34, 278)&0.9$\times 10^{-2}$&3.4&6.6&10.4&11.7\\
				\noalign{\smallskip}\hline\noalign{\smallskip}
				\multirow{4}{*}{\shortstack{\acs{map} \\(Forward\\greedy search)}} 
				&(11, -1)&0.9&3.6&7.5&10.5&12.2\\
				&(16, -1)&1.4&3.9&7.8&11.2&12.7\\
				&(21, -1)&2.1&4.1&8.1&11.5&12.9\\
				&(34, -1)&3.9&4.0&8.2&12.6&15.1\\
				&(34, 278)&23.5&3.9&8.3&12.6&15.1\\
				\noalign{\smallskip}\hline\noalign{\smallskip}
				\multirow{4}{*}{\shortstack{\acs{knn} \\ (Forward\\greedy search)} } 
				&(11, -1)&0.2$\times 10^{-2}$&4.1&7.4&11.3&12.9\\
				&(16, -1)&0.3$\times 10^{-2}$&4.4&7.6&12.4&15.6\\
				&(21, -1)&0.3$\times 10^{-2}$&4.4&8.1&12.6&18.0\\
				&(34, -1)&0.4$\times 10^{-2}$&4.7&8.8&13.5&20.9\\
				&(34, 278)&0.9$\times 10^{-2}$&3.3&6.4&10.7&11.2\\
				\noalign{\smallskip}\hline\noalign{\smallskip}
				\multirow{4}{*}{\shortstack{\acs{map} \\(\colorRev{\acs{foba}})}} 
				&(11, -1)&1.0&3.6&7.4&10.9&13.1\\
				&(16, -1)&1.7&3.8&7.8&11.7&13.6\\
				&(21, -1)&2.5&3.9&8.0&11.9&14.1\\
				&(34, -1)&5.1&4.0&8.3&12.8&15.6\\
				&(34, 278)&23.5&4.3&8.3&12.9&15.1\\
				\noalign{\smallskip}\hline\noalign{\smallskip}
				\multirow{4}{*}{\shortstack{\acs{knn} \\ (\colorRev{\acs{foba}})} } 
				&(11, -1)&0.2$\times 10^{-2}$&4.1&7.3&11.6&14.4\\
				&(16, -1)&0.3$\times 10^{-2}$&4.3&7.9&12.2&17.0\\
				&(21, -1)&0.3$\times 10^{-2}$&4.7&8.7&12.5&19.2\\
				&(34, -1)&0.4$\times 10^{-2}$&4.8&8.6&12.6&18.5\\
				&(34, 278)&0.9$\times 10^{-2}$&3.3&6.6&10.2&10.9\\
				\noalign{\smallskip}\hline\noalign{\smallskip}
		\end{tabular}}
	\end{table*}

	
	\section{Conclusion}\label{sec:futureWork}
	
	We proposed herein an approach to fingerprinting-based indoor positioning using opportunistically measured WLAN and BLE signals as the features for coordinate estimation. The main contributions are proposals to reduce data storage requirements and computational complexity in terms of processing time by segmentation of the entire  \acf{roi} into subregions, identification of a few candidate subregions during the online positioning stage, and use of a selected subset of relevant features instead of all available features for position estimation. Subregion selection is based on a modified Jaccard index quantifying the similarity between the features obtained by the user and those available within the \acs{rfm}. Feature selection is based on an adaptive forward-backward greedy search yielding a pre-computed set of relevant features for each subregion. The reduction of computational complexity is obtained both from the reduction of the number of candidate locations needed to analyze during online positioning and from the reduction of the number of features to be compared. 
	
	\colorText{The experimental results corroborated the claim of reduced complexity while indicating that the positioning accuracy is hardly reduced by subregion and feature selection for the small \acs{roi} and even improves for the large \acs{roi}. For both investigated \acsp{roi}, the time required for the position estimation in the online stage was reduced by a factor of about 10 when using the selected relevant features within 11 selected subregions instead of using all features and searching over the entire  \acs{roi}. For the small \acs{roi} (\ie \acs{roi} 1), the increment of the $ \texSym{90^{th}} $ percentile errors (CE90) is 7.5\% (\ie 8.6~m instead of 8.0~m).  In the large \acs{roi} (\ie \acs{roi} 2),  the positioning accuracy using \acs{map} reduces from 9.8~m to 7.4~m for \acs{wlan} signals and from 12.3~m  to 10.9~m for \acs{ble}. For \acs{knn}, the positioning accuracy does not change with subregion and feature selection.} Given a fixed number of candidate subregions and a fixed, low number of features the computational burden of the entire algorithm is almost independent of the size of the entire \acs{roi} and of the number of available features across the \acs{roi}.
	
	\colorText{Future research will concentrate on investigating the role of subregion definition (shape, orientation, homogeneity) and possible benefit of optimization, on taking into account} a user motion model during subregion selection, on handling temporal changes of the reference fingerprinting map, and on fully integrating different types of features for improving the positioning accuracy. 
	
	\section*{Acknowledgment}
	{The China Scholarship Council  (CSC) financially supports the first author's doctoral research. \colorText{We thank to anonymous reviewers for their useful comments and questions which helped significantly improving the paper.} }

\bibliographystyle{chicago}
\bibliography{lbs_2018}

\end{document}

%% file: cz_acronyms.tex
\acrodef{vc}[VC]{Vapnik-Chervonenkis}
\acrodef{knn}[$ k $NN]{$ k $ nearest neighbors}
\acrodef{lbs}[LBS]{location-based service}
\acrodef{ilbs}[ILBS]{indoor location-based service}
\acrodef{fips}[FIPS]{fingerprinting-based indoor positioning system}
\acrodef{fbp}[FbP]{fingerprinting-based positioning}
\acrodef{gnss}[GNSS]{global navigation satellite system}
\acrodef{ips}[IPS]{indoor positioning system}
\acrodef{rfid}[RFID]{radio frequency identification}
\acrodef{uwb}[UWB]{ultra wideband}
\acrodef{wlan}[WLAN]{wireless local area network}
\acrodef{rss}[RSS]{received signal strength}
\acrodef{ap}[AP]{access point}
\acrodef{roi}[RoI]{region of interest}
\acrodef{lasso}[LASSO]{least absolute shrinkage and selection operator}
\acrodef{rfm}[RFM]{reference fingerprint map}
\acrodef{map}[MAP]{maximum a posteriori}
\acrodef{cpa}[CPA]{cumulative positioning accuracy}
\acrodef{mse}[MSE]{mean squared error}
\acrodef{tp}[TP]{test position}
\acrodef{wrt}[w.r.t.]{with respect to}
\acrodef{mac}[MAC]{media access control}
\acrodef{imu}[IMU]{inertial measurement unit}
\acrodef{foba}[AFBGS]{adaptive forward-backward greedy search}
\acrodef{rp}[RP]{reference point}
\acrodef{mji}[MJI]{modified Jaccard index}
\acrodef{ble}[BLE]{Bluetooth low energy}
\acrodef{oil}[OIL]{organic indoor localization}
\acrodef{will}[WILL]{wireless indoor localization without site survey}
\acrodef{hiwl}[HIWL]{hidden Markov model-based indoor wireless localization}
\acrodef{svm}[SVM]{support vector machine}
\acrodef{lda}[LDA]{linear discriminant analysis}
\acrodef{sop}[SoP]{signals of opportunity}
\acrodef{ks}[KS]{Kolmogorov-Smirnov}
\acrodef{ecdf}[ECDF]{empirical cumulative distribution function}
\acrodef{ce}[CE]{circular error}

%% file: cz_command.tex
\newcommand{\todo}[1]{\hl{\textit{Tbd: #1}}}
\newcommand{\secPref}{Section }
\newcommand{\figPref}{Figure }
\newcommand{\tabPref}{Table }
\newcommand{\eg}{e.g., }
\newcommand{\ie}{i.e. }
\newcommand{\etal}{et al. }
\newcommand{\rhl}[1]{\textcolor{red}{\hl{#1}}}
\newcommand{\colorText}[1]{{#1}}
\newcommand{\colorRev}[1]{{#1}}
\newcommand{\convSetSym}[1]{\varmathbb{#1}}
\newcommand{\setSym}[1]{\mathbbm{#1}}
\newcommand{\vecSym}[1]{\mathbf{#1}}
\newcommand{\funSym}[1]{\mathpzc{#1}}
\newcommand{\textSym}[1]{\mathrm{#1}}
\newcommand{\mapSym}{\mapsto}
\newcommand{\texSym}[1]{\mathrm{#1}}
\newcommand{\intSet}[1]{\{1, 2, \cdots, #1\}}
\newcommand{\setSymScript}[3]{\setSym{#1}_{#2}^{\texSym{#3}}}
\newcommand{\vecSymScript}[3]{\vecSym{#1}_{#2}^{\texSym{#3}}}
\newcommand{\norSymScript}[3]{#1_{#2}^{\texSym{#3}}}
\newcommand{\compComp}[1]{\mathcal{O}(#1)}
\newcommand\blankpage{%
	\null
	\thispagestyle{empty}%
	\addtocounter{page}{-1}%
	\newpage}